\documentclass[aps,prd,twocolumn,nofootinbib,superscriptaddress,floatfix]{revtex4-1}
\usepackage[colorlinks=false, linkcolor=black, citecolor=black]{hyperref}
\usepackage{graphicx}
\usepackage{subfig}
\usepackage{amstext,amsmath,amssymb,amsfonts,amsthm}
\usepackage[format=plain,justification=centerlast]{caption}

\def\app#1{{Appendix~\ref{#1}}}
\def\sec#1{{Section~\ref{#1}}}
\def\eq#1{{Eq.~(\ref{#1})}}

\def\fig#1{{Fig.~\ref{#1}}}

\def\dd{\mathrm{d}}
\def\be{\begin{equation}}
\def\ee{\end{equation}}
\def\bes{\begin{eqnarray}}
\def\ees{\end{eqnarray}}
\def\ba{\begin{align}}
\def\ena{\end{align}}
\def\bwt{\begin{widetext}}
\def\ewt{\end{widetext}}

\def\bra{\langle}
\def\ket{\rangle}
\def\f{\frac}

\def\pp{\partial}
\def\nn{\nonumber}

\def\tbf#1{{\textbf{#1}}}
\def\inf{\mathrm{inf}}

\def\calS{{\mathcal S}}

\begin{document}

\title{Evolution of quantum field, particle content and classicality in the three stage universe}
\date{\today}
\author{Suprit Singh}
\email{suprit@iucaa.ernet.in}
\affiliation{Inter-University Centre for Astronomy and Astrophysics, Ganeshkhind, Pune 411 007, India.}
\author{Sujoy Kumar Modak}
\email{sujoy@iucaa.ernet.in}
\email{sujoy.kumar@correo.nucleares.unam.mx}
\affiliation{Instituto de Ciencias Nucleares, Universidad Nacional Aut\'{o}noma de M\'{e}xico, Apartado Postal 70-543, Distrito Federal, 04510, M\'{e}xico}
\affiliation{Inter-University Centre for Astronomy and Astrophysics, Ganeshkhind, Pune 411 007, India.}
\author{T. Padmanabhan}
\email{paddy@iucaa.ernet.in}
\affiliation{Inter-University Centre for Astronomy and Astrophysics, Ganeshkhind, Pune 411 007, India.}

\begin{abstract}
We study the evolution of a quantum scalar field in a toy universe which has three stages of evolution, viz.,
(i) an early (inflationary) de Sitter phase  (ii) radiation dominated phase  and (iii) late-time (cosmological constant dominated) de Sitter phase. Using Schr\"odinger picture, the  scalar field equations are solved separately for the three stages and matched at the transition points. The  boundary conditions are chosen so that field modes in the early de Sitter evolves from Bunch-Davies vacuum state. We determine the (time-dependent) particle content of this quantum state  for the entire evolution of the universe and describe the various features both numerically and  analytically.  We also describe the quantum to classical transition  in terms of a {\it classicality parameter} which tracks the particle creation and its effect on phase space correlation of the quantum field.  
\end{abstract}

\maketitle

\section{Introduction}

It is well known that quantum vacuum becomes unstable and gives rise to particle production when the external sources are sufficiently strong. Some familiar examples of this phenomena are the particle creation by electric fields (Schwinger effect) \cite{schwinger},  by black holes  \cite{hawk} and in the expanding universe \cite{early}. 
Studies on quantum field theory in curved spacetime \cite{early}-\cite{others} have provided more general examples of particle creation in different contexts. 

When the external sources can be switched off asymptotically, leading to free field theory at very early and late times, one can define \textit{in} and \textit{out} vacua in the asymptotic regimes and study particle creation in a reasonably unambiguous manner.  However, many interesting cases  in the study of quantum field theory in curved spacetime do not allow us the luxury of asymptotic vacua. Cosmological particle creation is one such example. In this context, one can certainly calculate the time evolution of a given quantum state in Heisenberg or Schr\"odinger picture in an unambiguous manner. But interpreting the \textit{particle content} of this quantum state at any given time is fraught with ambiguities.  It is generally recognized that one cannot resolve these ambiguities by any unique procedure which is applicable in  all contexts. The best one could  do is to introduce different constructs which could probe different aspects of physics in the expanding background and develop an intuitive feel for the various phenomena. 

Another question, closely related to particle content, is the notion of classicality. A quantum field in an external background might have features which possess nearly classical description in certain contexts. For example, it is believed that fluctuations of  a scalar field, which were purely quantum mechanical in origin in the early stages of the inflationary phase, allow a description as purely classical stochastic fluctuation at the late stages. Much of the current paradigm in cosmology \cite{brand}-\cite{othr} pre-supposes such a notion of quantum to classical transition. Once again, it is not possible to quantify the degree of classicality of a field in a unique and all encompassing manner. The best we could do is to come up with measures of classicality and see how best they work in different contexts.

It would be useful, intuitively clear and somewhat economical if we could come up with constructs which simultaneously give a handle on the degree of classicality of the field and its particle content. In fact, such a criterion will sharpen our intuitive idea that well defined notion of particles will exist if and only if the degree of classicality is high, while the notion of particles will be drowned in the sea of quantum fluctuations when the degree of classicality is low. Such a procedure and a fairly comprehensive methodology was proposed sometime back in a series of papers \cite{gaurang2007,gaurang2008}.
This work exploits the fact that the quantum theory of a minimally coupled scalar field in a Friedmann spacetime can be reduced to the study of a time dependent harmonic oscillator. 

The Schr\"odinger equation for the time dependent harmonic oscillator admits \textit{form-invariant} Gaussian states as solutions (this is well-known in the literature, see for example~\cite{sriram}), which --- in turn --- allows us to define {\it instantaneous} particle content of the state. Particles are defined in terms of instantaneous eigenstates specified at each moment. This approach has proved to be quite successful in dealing with time dependent particle content in various cases, like for example--- massless minimally coupled scalar field in de Sitter and radiation dominated Friedmann spacetime \cite{gaurang2007}; complex scalar field in a constant electric field \cite{gaurang2007} and time dependent electric field \cite{gaurang2008} backgrounds. The quantum to classical transition is usually discussed in terms of Wigner function which links the wavefunction that appears in Schr\"odinger's equation to a probability distribution in phase space. However, it was found \cite{gaurang2007} that the peaking of the Wigner function on the classical phase space trajectory is independent of particle content in several contexts and hence makes the definition of classicality in terms of Wigner function somewhat less useful. To counter this, an additional measure of phase space correlation (called the \textit{classicality parameter}) was proposed in \cite{gaurang2007} which maintains our intuitive link between degree of classicality and particle content. When there is no particle creation this parameter is zero and in the presence of strong particle creation its modulus saturates at the maximum value of unity.
 
Motivated by these studies, we analyse in this paper the particle content and classicality of a quantum field, by considering the entire evolutionary history of our universe. Because the matter dominated phase in our universe has lasted only for about 4 decades of expansion, while the radiation dominated phase has lasted for nearly 24 decades of expansion, we approximate the evolutionary history of the universe as made up of just three stages --- the early (inflationary) de Sitter phase,  a radiation dominated phase and  late-time de Sitter phase characterizing the current accelerated expansion of the universe. The time-dependent minimally coupled scalar field equations are solved, separately, in these three stages. The field solutions as well as the scale factors corresponding different regions are then matched at two transition points (i.e., de Sitter to radiation and radiation to de Sitter transitions). These allow us to discuss issues like time dependent particle content, emergence of classicality for comoving case in an integrated manner.

This paper is organized as follows. In \sec{sec:schrdyn} we provide a brief summary of the formalism developed in \cite{gaurang2007}, mainly focusing the  particle content and degree of classicality. \sec{sec:3phases} is devoted to the construction of the background spacetimes that we call our three stage Universe. Here we solve the time dependent scalar field equations and mode solutions are matched at transition points. The boundary condition is chosen in such a way that all modes evolve from Bunch-Davies vacuum state defined in the initial inflationary de Sitter phase. At any later time these modes evolves non-adiabatically and particle creation takes place in all three stages. \sec{sec:pcc} includes a detailed numerical as well as analytical description of the particle content as a function of scale factor and the behavior of classicality parameter in various regimes. We present the conclusions in \sec{sec:dc}.      

\section{Schr\"{o}dinger dynamics of a Quantum Field}
\label{sec:schrdyn}

To set the stage, we will begin by summarizing the  formalism developed in~\cite{gaurang2007} to study quantum fields in an expanding universe and recall the key ideas related to definition of states, particle creation and classicality. We will not provide detailed motivation of these ideas here; the interested reader may find more details in~\cite{gaurang2007}. 

It is well-known that the dynamics of free fields in the Friedmann background can be reduced to that of decoupled, time dependent, harmonic oscillators which can be quantized in Schr\"odinger picture. We consider a massless minimally coupled scalar field in the spatially flat Friedmann background  
\bes
\label{metric}
ds^2 &=& dt^2 - a^2(t)d\tbf x^2 \nn\\
&=& a^2(\eta)(d\eta^2  - d\tbf x^2)
\ees
where the conformal time is defined by $\eta\equiv\int \dd t \, a^{-1}$. The action for the field is then given by
\bes
\label{actn1}
\calS[\Phi(\eta,\tbf x)] &=& \int \dd^4 x\,\sqrt{-g}\,\pp_a\Phi \pp^a\Phi \nn\\
&=&\f{1}{2}\int \dd^3 \tbf x\,\int\dd\eta\, a^2(\eta)\left(\pp^2_\eta\Phi - \pp^2_{\tbf x}\Phi\right).
\ees
Due to the translational invariance of the metric in \eq{metric}, one can  decompose the field into independent Fourier modes as 
\be
\Phi(\eta,\tbf x) = \int \f{\dd^3 k}{(2\pi)^3}\,\xi_{\tbf k}(\eta) e^{i\tbf{k}\cdot\tbf{x}}
\ee
Since $\Phi$ is real, this implies $\xi_{\tbf k} = \xi_{-\tbf k}^*$ for the  complex $\xi_{\tbf k}$. This constraint essentially halves the degrees of freedom in $\xi_{\tbf k}$, so that in terms of a single real variable $\phi_{\tbf k}$, one can express the action in \eq{actn1} as
\be
\label{reducedactn}
\calS[\phi_{\tbf k}(\eta)] =\f{1}{2}\int \dd^3 \tbf k\,\int\dd\eta\, a^2(\eta)\left(\dot{\phi}_{\tbf k}^2 - k^2\phi_{\tbf k}^2\right)
\ee
where the dot implies the derivative with respect to $\eta$ and $k = |\tbf k|$. The field system thus gets reduced to a bunch of decoupled harmonic oscillators in the Fourier domain with time-dependent mass, $a^2(\eta)$ and frequencies, $k$. We can now now use the fact that this Schr\"{o}dingier equation admits  time-dependent, form-invariant, Gaussian states with vanishing mean given by:
\bes
\label{wavefn}
\psi(\phi_{\tbf k},\eta) &=& N \exp\left[-\alpha_k(\eta)\phi_{\tbf k}^2\right] \nn\\
&=& N \exp\left[-\f{a^2(\eta) k}{2}\left(\f{1- z_k}{1+z_k}\right)\phi_{\tbf k}^2\right]
\ees
The time evolution of the  wave function is now given in terms of that of the functions $\alpha_k(\eta)$ and $z_k(\eta)$ which satisfy the equations:
\be
\label{eqnforalpha}
\dot{\alpha}_k = \f{2\alpha_k^2}{a^2} - \f{1}{2} a^2 k
\ee
and 
\be
\dot{z}_k + 2ikz_k + \left(\f{\dot{a}}{a}\right)(z_k^2 -1) = 0
\ee
Here we use the notation and terminology introduced in \cite{gaurang2007} in which $z_k$, called the {\it excitation parameter},  measures the deviation of $\alpha_k$ from the adiabatic value. (The functions $\alpha_k$ and $z_k$ depend only on the modulus of \tbf k and hence the subscripts are not in boldface.) We thus need to solve for $\alpha_k$ or $z_k$ to infer the quantum evolution of the system and related characteristics. The non-linear first order equations can be related to the second-order \textit{linear} differential equation by introducing  another function $\mu_k(\eta)$, defined through the relation $\alpha_k = - (ia^2/2) (\dot{\mu}_k/\mu_k)$, which satisfies:   
\be
\label{eqnformu}
\ddot{\mu}_k + 2\left(\f{\dot{a}}{a}\right)\dot{\mu}_k + k^2 \mu_k = 0
\ee
This is the same as the field equation for $\phi_\tbf k$ resulting from varying the action in \eq{reducedactn}. As for the function $z_k$, it is related to $\mu_k$ by:
\be
\label{zmu}
z_k = \left(\f{k\mu_k +  i \dot{\mu}_k}{k\mu_k -  i \dot{\mu}_k}\right)
\ee
Thus it suffices to solve for $\mu_k$ given the boundary conditions to determine the quantum evolution of the system. 

Since $\mu_k$ satisfies the second order linear differential equation, it will have two linearly independent solutions and thus we can write, in general, $\mu_k(\eta) = \mathcal{A}_k s_k(\eta) + \mathcal{B}_k  s_k^*(\eta)$. But  $z_k$ and $\alpha_k$ depends only on  the ratio $\dot{\mu}_k/{\mu_k}$ so that the overall normalization of $\mu_k$ is irrelevant and the evolution only depends on the ration $\mathcal{R}_k = \mathcal{B}_k/\mathcal{A}_k$ for given initial conditions. 

The initial conditions are set such that the state described by the wave function in \eq{wavefn} is a ground state with zero particle content at some time say $\eta = \eta_i$ when $a (\eta_i) = a_i $. (It is often convenient to use the scale factor itself as a  time variable with the  replacements $d/d\eta \rightarrow \dot{a}\,d/da$ etc.) Then, the initial condition of the wave function at $a=a_i$ being the ground state wave function of an harmonic oscillator demands,
\be
\alpha_k(a_i) = \f{a^2_i k}{2}
\ee
or equivalently $z_k(a_i) = 0$,  implying
\be
\label{cndtnmu}
\left(\f{\dot{a}}{\mu_k}\left.\f{\dd{\mu}_k}{\dd a}\right)\right|_{a_i} = i k
\ee
which in turn determines $\mathcal{R}_k$ thereby fixing the state. As the system evolves, we are interested in the two specific quantities: the particle content of the state at any time and the degree of classicality of the state. Both of these were discussed in detail in \cite{gaurang2007} and we will just summarize the motivation here and adopt the ideas: Our initial condition implies that the state begins as a ground state at $a=a_i$ but at any later time will be different from the instantaneous ground state. To quantify the instantaneous particle content of this state it is then reasonable to compare it with the \emph{instantaneous} eigenstates at every instant obtained by adiabatically evolving the  the eigenstates at some initial epoch. Since $\psi$ is an even function, the overlap is non-zero only with even eigenstates. One can calculate this overlap (for details the reader is referred to \cite{gaurang2007}) with the eigenstates to get time-dependent probability distribution of transitions and using it, the mean number of quanta in the state at any time can be computed.
This particle content is given in terms of $z_k$ as
\be
\label{n}
\bra n_k\ket = \f{|z_k|^2}{1-|z_k|^2}
\ee 
Note that being time-dependent and related to transitions within the instantaneous eigenstates, we do \textit{not} expect $\bra n_k\ket$ to be monotonic in general. The mean occupation number can go up and down and hence should not be taken to be the `particle' content in the \emph{classical} sense since it can be accompanied by fluctuating \emph{quantum} noise when the system is far away from classicality. 

The degree of classicality of the state brings us to the next quantity of interest, the \emph{classicality} parameter $\mathcal{C}_k$, which is the measure of phase space correlations of the system. Classicality is usually  quantified by the use of Wigner distributions and  is inferred from the peaking of the distribution on the corresponding classical trajectory. However, it can be shown \cite{gaurang2007} that the naive reliance on Wigner function can lead to ambiguities. Hence a new correlation function, $\mathcal{C}_k$ was introduced in \cite{gaurang2007} as a more robust measure to quantify classicality and it is found to be in excellent agreement with our intuitive ideas in many cases. This quantity is given by
\be
\mathcal{C}_k = \f{\mathcal{J}_k\sigma_k^2}{\sqrt{1+(\mathcal{J}_k\sigma_k^2)^2}}
\ee  
where $\mathcal{J}_k$ and $\sigma_k$ are the parameters of Wigner function defined in the $\phi_\tbf k - \pi_\tbf k$ phase space of the oscillator for the Gaussian state in \eq{wavefn}
\be
\mathcal{W}(\phi_{\tbf k},\pi_{\tbf k},\eta) = \f{1}{\pi}\exp\left[-\f{\phi^2_{\tbf k}}{\sigma_k^2} - \sigma_k^2(\pi_\tbf k - \mathcal{J}_k \phi_\tbf k)^2\right].
\ee
In terms of $z_k$ we have,
\be
\label{cp}
\mathcal{C}_k = \f{2\,\mathrm{Im}(z_k)}{1 - |z_k|^2}.
\ee
The vanishing of $\mathcal{C}_k$ implies $\mathcal{J}_k = 0$ and Wigner distribution is then an uncorrelated product of gaussians in $\phi_\tbf k$ and $\pi_\tbf k$ which is the case for the ground state which itself is gaussian, set up as the initial condition. Otherwise the classicality parameter is confined to the interval $[-1,1]$ when the Wigner function becomes correlated. The particle creation and classicality of a state are strongly associated with each other; as we shall see the notion of particles become well-defined when the degree of classicality is high and vice-versa. With the structure in place, we shall get on with our study of these aspects in the cosmological context.


\section{The Three Stage Universe}
\label{sec:3phases}

We consider a three stage universe consisting of an initial inflationary de Sitter phase characterized by the Hubble parameter, $H_\inf$ which evolves into a radiation dominated phase and ends up in a late-time de Sitter phase dominated by a cosmological constant $\Lambda$. Equivalently the final phase can be characterized by the Hubble parameter, $H_\Lambda$ with $H_\Lambda^2=\Lambda/3$. (This model ignores the matter dominated phase for simplicity, which can be justified by the fact that --- in our universe ---  the matter domination lasts only for about 4 decades while radiation domination lasted for about 24 decades.) The scale factor for such a three-stage universe can be taken to be:
\begin{figure}[t!]
\includegraphics[scale=0.50,width=0.49\textwidth]{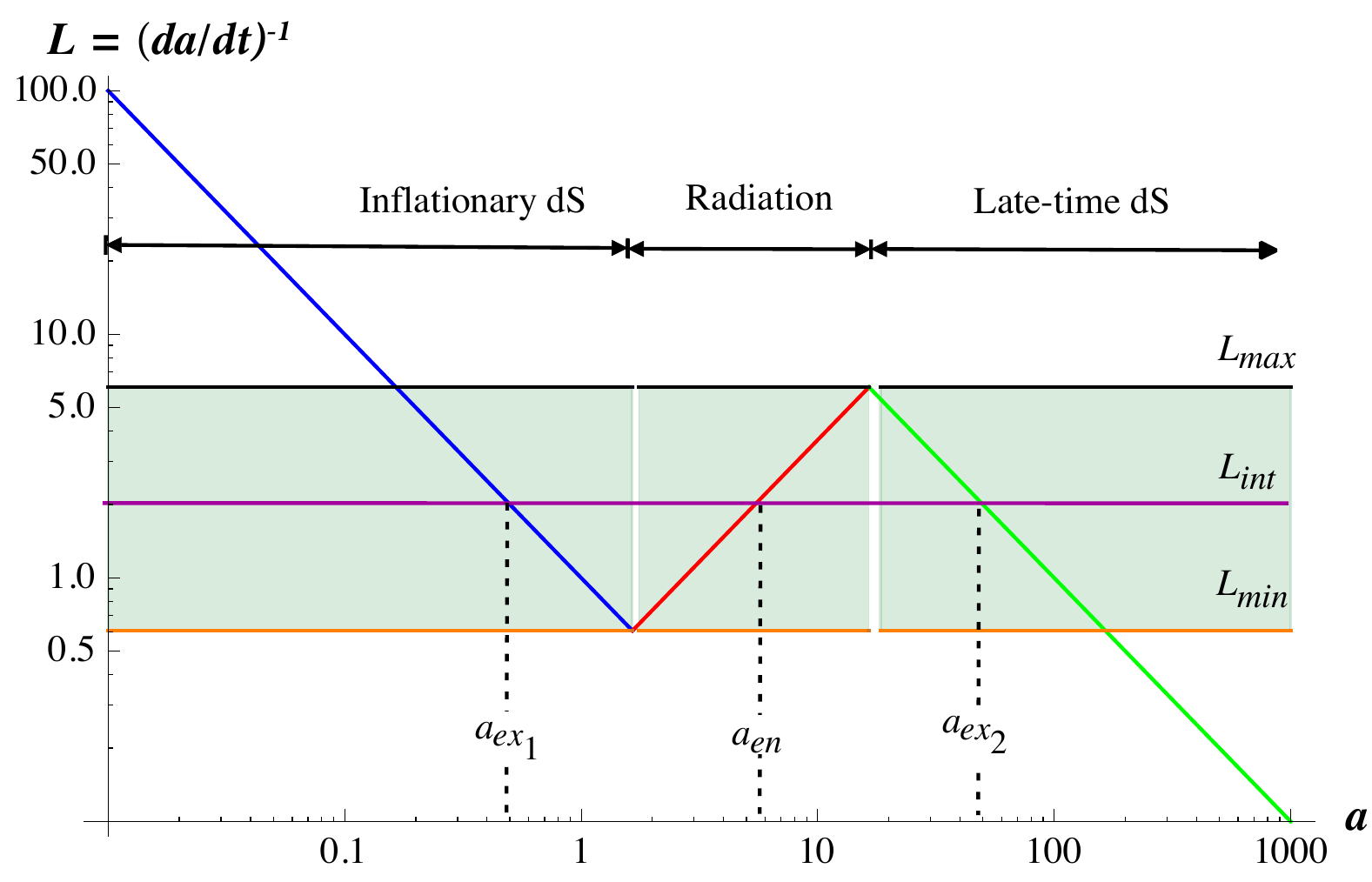}
\caption{The three stage universe. Evolution of comoving Hubble radius, $L = (da/dt)^{-1}$ with the scale factor $a$ for $\epsilon = 0.01$ is shown in the inflationary phase (decreasing), radiation-dominated phase (increasing) and in late-time de Sitter phase (decreasing) where the lines have slopes $\pm 1$ in the logarithmic plot. We have two characteristic length scales, $L_{max}$ and $L_{min}$ forming a band such that any length scale within the band has three transition points where it goes super-Hubble, sub-Hubble and finally super-Hubble again.}
\label{comovuniv}
\end{figure} 

\begin{multline}
\label{aoft}
a(t) =
\left
\{
\begin{array}{ll}
 e^{H_\inf t}  & \hspace{10pt}t \le t_r \\
        (2H_\inf e)^{1/2}\, t^{1/2} &  \hspace{10pt} t_r\le t \le t_{\Lambda}\\
        (H_\inf/H_\Lambda)^{1/2}e^{H_\Lambda t}  &\hspace{10pt}  t \ge t_{\Lambda}
\end{array}
\right.
\end{multline}
where $t_r = (2H_\inf)^{-1}$ and $t_{\Lambda} = (2H_\Lambda)^{-1}$ are the comoving times at the two respective transitions and we ensured that the scale factor and its derivative are continuous at the transition points. Further, we shall take  $H_\Lambda = \epsilon H_\inf = \epsilon H$ with $\epsilon\ll 1$, since for our real universe, $\epsilon\sim10^{-54}$. However,  for visual clarity of figures, we will use sufficiently small  values for $\epsilon$ to discuss the effects and comment about the cosmological case later. 
We can re-express the scale factor in terms of the conformal time  as 
\begin{multline}
\label{aofeta}
a(\eta) =
\left
\{
\begin{array}{ll}
 \left[(\eta_r - \eta)H+ e^{-1/2}\right]^{-1}  & \hspace{10pt}\eta \le \eta_r \\
 (e/\epsilon)^{1/2}+ He (\eta - \eta_\Lambda) &  \hspace{10pt} \eta_r \le \eta \le \eta_\Lambda\\
  \left[(\epsilon/e)^{1/2} - \epsilon H (\eta - \eta_\Lambda)\right]^{-1} & \hspace{10pt}  \eta \ge \eta_\Lambda
        \end{array}
\right.
\end{multline}
with $(\eta_f - \eta_\Lambda) = (He^{1/2})^{-1}(\epsilon^{-1/2} -1)$. But it is more convenient to use  $a$ itself as the time-variable since it gives a better conceptual understanding of the length scales involved. Note that $a_r = a (t_r) = e^{1/2}$ and $a_\Lambda = a(t_\Lambda) = (e/\epsilon)^{1/2}$ at the transition points with our choice of normalization. The comoving Hubble radius, in terms of the scale factor, is given by
\begin{multline}
\label{comovlength}
   L(a) = (\dot{a})^{-1} = \left\{
     \begin{array}{ll}
        (H a)^{-1} &\hspace{7pt}a \le e^{1/2} \\
        (a/ H e) &  \hspace{7pt} e^{1/2}\le a \le (e/\epsilon)^{1/2}\\
        (a\epsilon H)^{-1}  &\hspace{7pt}   a\ge (e/\epsilon)^{1/2}
     \end{array}
   \right.
\end{multline}  
The above scheme is pictorially depicted in \fig{comovuniv}. In the logarithmic plot the lines have unit slope and are at $45$ degrees. The comoving Hubble radius shrinks during the initial inflationary de Sitter (blue) phase till $a_r = e^{1/2}$, when the radiation dominated phase starts and the comoving Hubble radius increases (red) until $a_\Lambda = (e/\epsilon)^{1/2}$ (which is decided solely by the value of $\epsilon$) the final de Sitter phase (green) sets in leading to the shrinkage again.

In such a universe, there exist two length scales, $L_{max} = 1/(H\epsilon^{1/2} e^{1/2})$ and $L_{min} = 1/(H e^{1/2})$ and a band in between which is special (see e.g. \cite{cosmin}). A wave mode characterized by the length scale larger than $L_{max}$, once exits the comoving Hubble radius in the initial de Sitter phase, will remain super-Hubble at all times. Similarly modes with wavelengths smaller than $L_{min}$ remains sub-Hubble till it exits the Hubble radius in the late-time de Sitter phase. Any wave mode characterized by the length scale, $L_{int}$ lying within the band encounters three transition points: it exits the Hubble radius at some $a_{ex_1}$ during initial de Sitter phase and goes super-Hubble, then enters the the Hubble radius at $a_{en}$ during the radiation dominated phase becoming sub-Hubble again and re-exits the Hubble radius in the final de Sitter phase at $a_{ex_2}$ becoming super-Hubble once again. While $L_{min}$ is independent of $\epsilon$, $L_{max}$ depends on it inversely which is expected, since $\epsilon$ determines the duration of the radiation dominated phase. This rich terrain, as we shall see, affects the particle creation aspects and classicality for a test scalar field in a non-trivial manner. 

We shall now apply the formalism of \sec{sec:schrdyn} in the toy cosmological model described above. This requires working out the evolution of the wave function in the three stages which, in turn, requires computing $\mu_k(a)$, with a given  initial condition in the first de Sitter phase and sewing it with the other two patches by demanding the continuity of the wave function and its derivative at the transition points (i.e., de Sitter $\rightarrow$ radiation $\rightarrow$ de Sitter transitions). We shall now turn to this task and describe the solution in the three stages.
 
\subsection{The inflationary phase}
\label{subsec:ds1}
With the scale factor specified for the case in \eq{aoft} or \eq{aofeta}, we can solve for $\mu_k(\eta)$ using \eq{eqnformu} and by inverting the function $a(\eta)$ we have,
\be
\mu_k^{(1)} (a) =  s_k (a) + \mathcal{R}_k s_k^*(a)
\ee
with
\be
s_k(a) =  \left(\f{1}{a} - \f{i H}{k}\right) \exp\left[\f{ik}{H}\left(1-\f{1}{a}\right)\right].
\ee
The constant $\mathcal{R}_k$ is determined by imposing the initial condition as per \eq{cndtnmu} at an initial epoch $a_i$, which gives:
\be
\mathcal{R}_k (a_i) = \left(1-\f{2ik}{Ha_i}\right)^{-1}\exp\left[\f{2ik}{H}\left(1 -  \f{1}{a_i}\right)\right]
\ee
Different choices of $a_i$ will correspond to different initial conditions. We will choose the state to be a ground state in the asymptotic past  that is, when $a_i\rightarrow 0$.  This can be achieved by choosing  $\mathcal{R}_k=0$  for all $k$. Thus we have
\be
\label{muk1}
\mu_k^{(1)} (a) =   \left(\f{1}{a} - \f{i H}{k}\right) \exp\left[\f{ik}{H}-\f{i k}{a H}\right]
\ee
This is the state evolved to an epoch $a$ from conventional Bunch-Davies vacuum ~\cite{bunchdavies} at the asymptotic past. At any finite time it is different from the instantaneous vacuum state and is a mixture of positive and negative frequency modes~\cite{suprit1302} with non-zero particle content. 
 
\subsection{Radiation dominated phase}
\label{subsec:radtn}
For the radiation dominated phase, $\mu_k(a)$ is given by
\be
\label{muk2}
\mu_k^{(2)} (a) =  \f{1}{a}\left(C_k e^{-ik a/eH } + D_k e^{ik a/eH}\right)
\ee
where $C_k$ and $D_k$ are determined by the matching conditions at $a =  a_r  = e^{1/2}$ 
\be
\label{match12}
\mu_k^{(1)} (a_r) = \mu_k^{(2)} (a_r);\hspace{10pt} \mu_k^{(1)'} (a_r) = \mu_k^{(2)'} (a_r)
\ee
where the prime denotes derivative with respect to $a$. This can be done analytically but the resulting expressions are not very illuminating.

\subsection{The late-time de Sitter phase}
\label{subsec:ds2}

The scale factor in the late-time de Sitter is related to the inflationary phase by the replacements $H\rightarrow \epsilon H$ and an overall scaling by $\epsilon^{-1/2}$. So we have 
\be
\label{muk3}
\mu_k^{(3)} (a) =  E_k \bar{s}_k (a) + F_k \bar{s}_k^*(a)
\ee
with
\be
\bar{s}_k(a) = \exp\left[\f{ik}{\epsilon H}\left(1-\f{1}{a}\right)\right] \left(\f{1}{a} - \f{i \epsilon H}{k}\right).
\ee
Again, $E_k$ and $F_k$ are determined by matching at $a =  a_\Lambda  = (e/\epsilon)^{1/2}$: 
\be
\label{match23}
\mu_k^{(2)} (a_\Lambda) = \mu_k^{(3)} (a_\Lambda);\hspace{10pt} \mu_k^{(2)'} (a_\Lambda) = \mu_k^{(3)'} (a_\Lambda).
\ee
With the help of \eq{match12} and \eq{match23} we now have $\mu_k(a)$ connected up for all the three stages. The next step is to compute the particle content and the classicality parameter in an integrated manner for the entire evolution history of the universe. This is done using \textit{Mathematica} for algebraic manipulations and the results are presented in the next section.  


\section{Particle content and Classicality}
\label{sec:pcc}
We now recall the method summarized in \sec{sec:schrdyn} to find the average particle number and classicality parameter given by \eq{n} and \eq{cp} respectively. The exact analytical results for these quantities are simple in the inflationary  phase, but gets algebraically unwieldy for the radiation phase and an impossibly complicated (having hundreds of terms!) in the final de Sitter phase. Hence we shall first present our results in \fig{epsilon1} and \fig{epsilon2} using results for a range of values of $\epsilon$ using algebraic manipulation software. Having described the exact results in this manner, we will provide approximate analytic expressions highlighting the behavior pattern in various limits in \sec{subsec:analyticlimits}.


\subsection{The numerical results}
As is evident from \eq{aofeta}, the smaller values of $\epsilon$ correspond to longer lifetime of the radiation dominated phase. Here we first consider toy universes with $\epsilon = 0.01$ and $\epsilon = 0.0001$ to capture the whole picture  which is numerically  difficult to do with the cosmologically relevant value of $\epsilon$. Note that, the sub and super-Hubble regions, as shown in \fig{comovuniv}, also provide an estimate for $k/H$. Modes with $k/H < e^{1/2} \epsilon^{1/2}$ are always super-Hubble once they exit the Hubble radius in the inflationary phase and modes with $k/H > e^{1/2}$ are always sub-Hubble until they exit the Hubble radius in the late-time de Sitter phase. Any other mode will exit, enter and re-exit the Hubble radius in different stages as shown in \fig{comovuniv}. For the set of expressions we come across here,  it is possible to normalize the the wave vector ($k$) with respect to $H$ so that one can set $H = 1$ for the numerical scheme. But it should be remembered that in this case $k$ really means $k/H$. We shall reinstate $H$ in the next subsection where we deal with analytical expressions. 

\begin{figure*}[t!]
\includegraphics[width=0.40\textwidth]{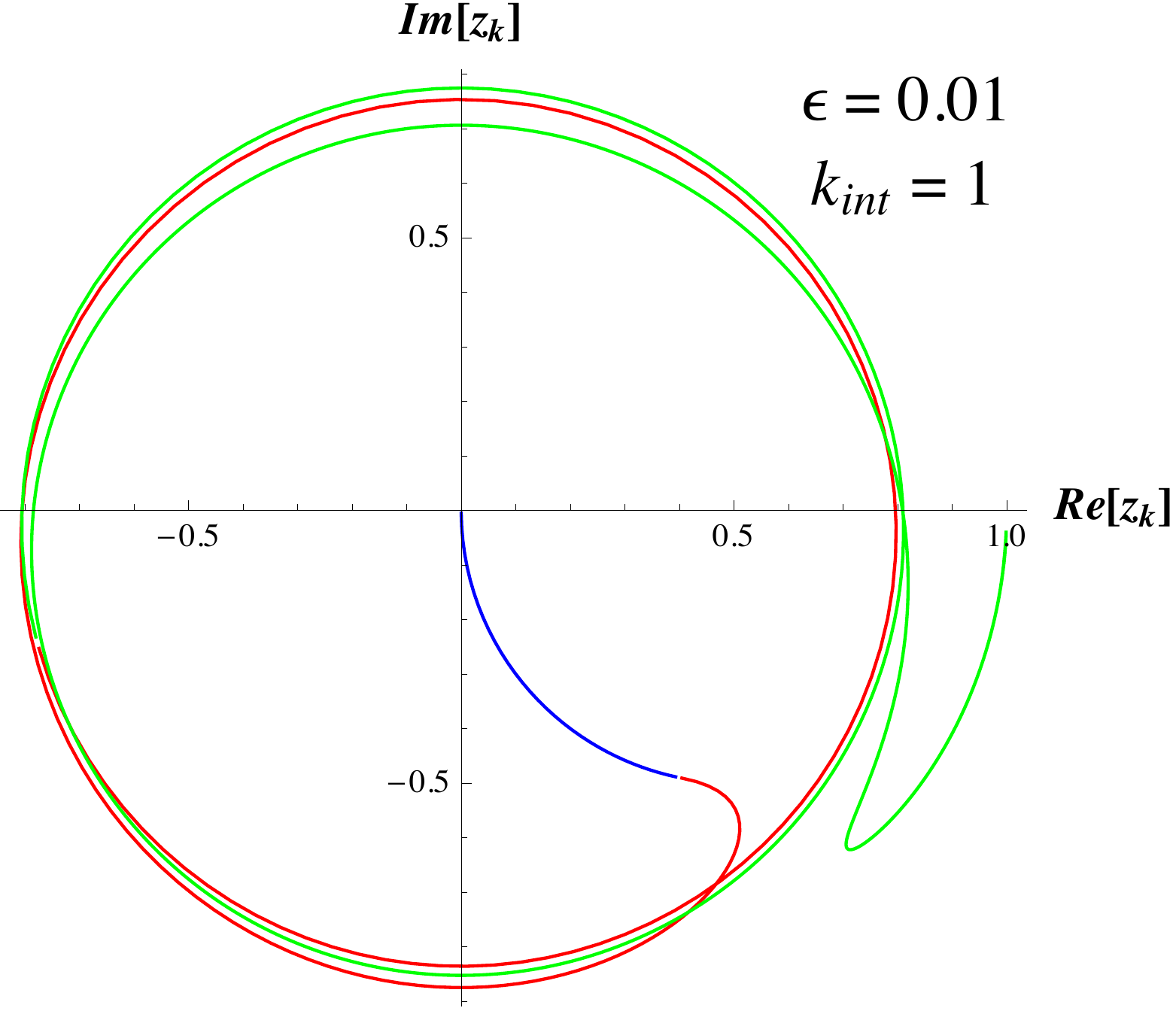}\hfill
\includegraphics[width=0.49\textwidth]{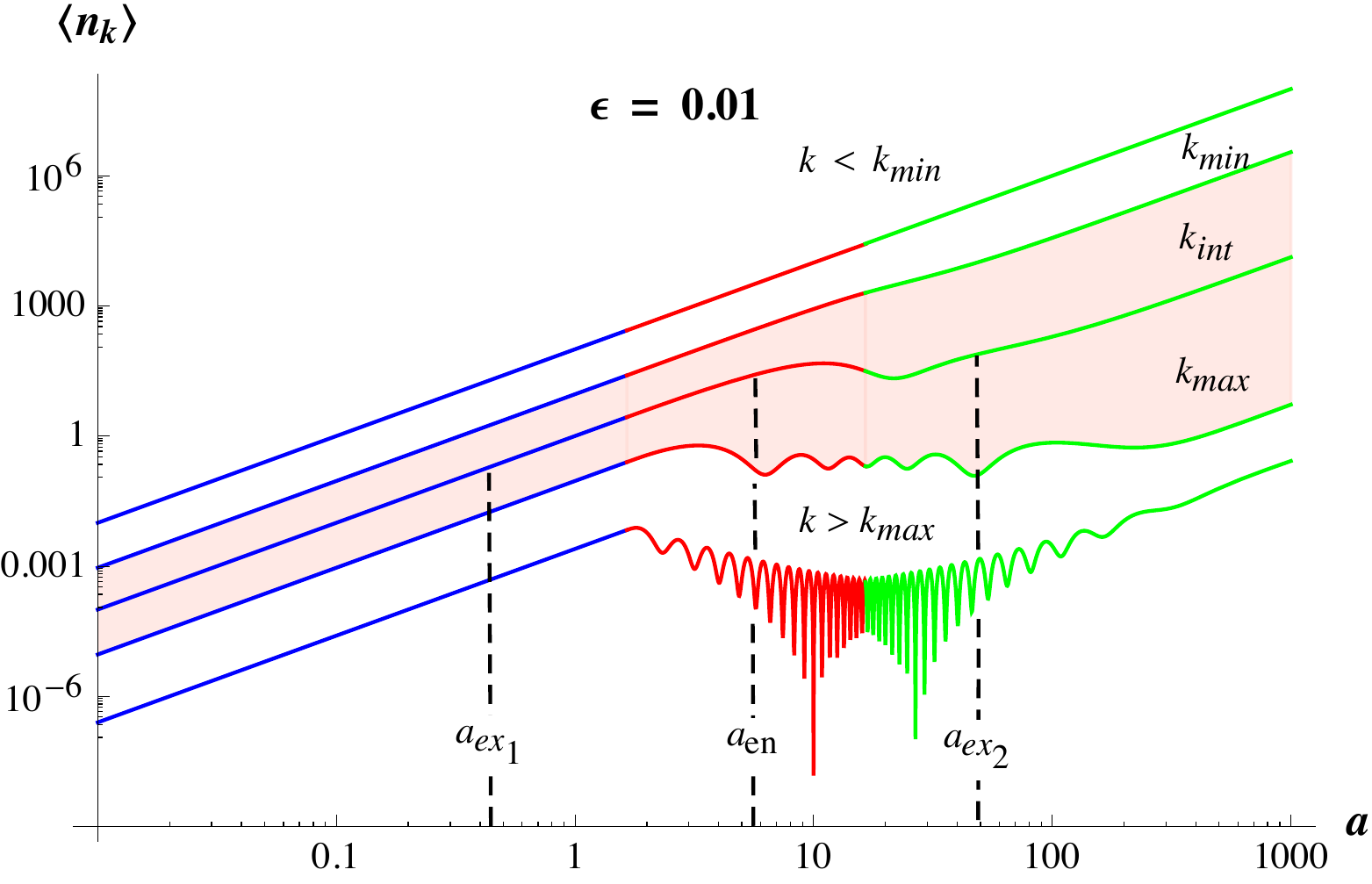}
\includegraphics[width=0.33\textwidth]{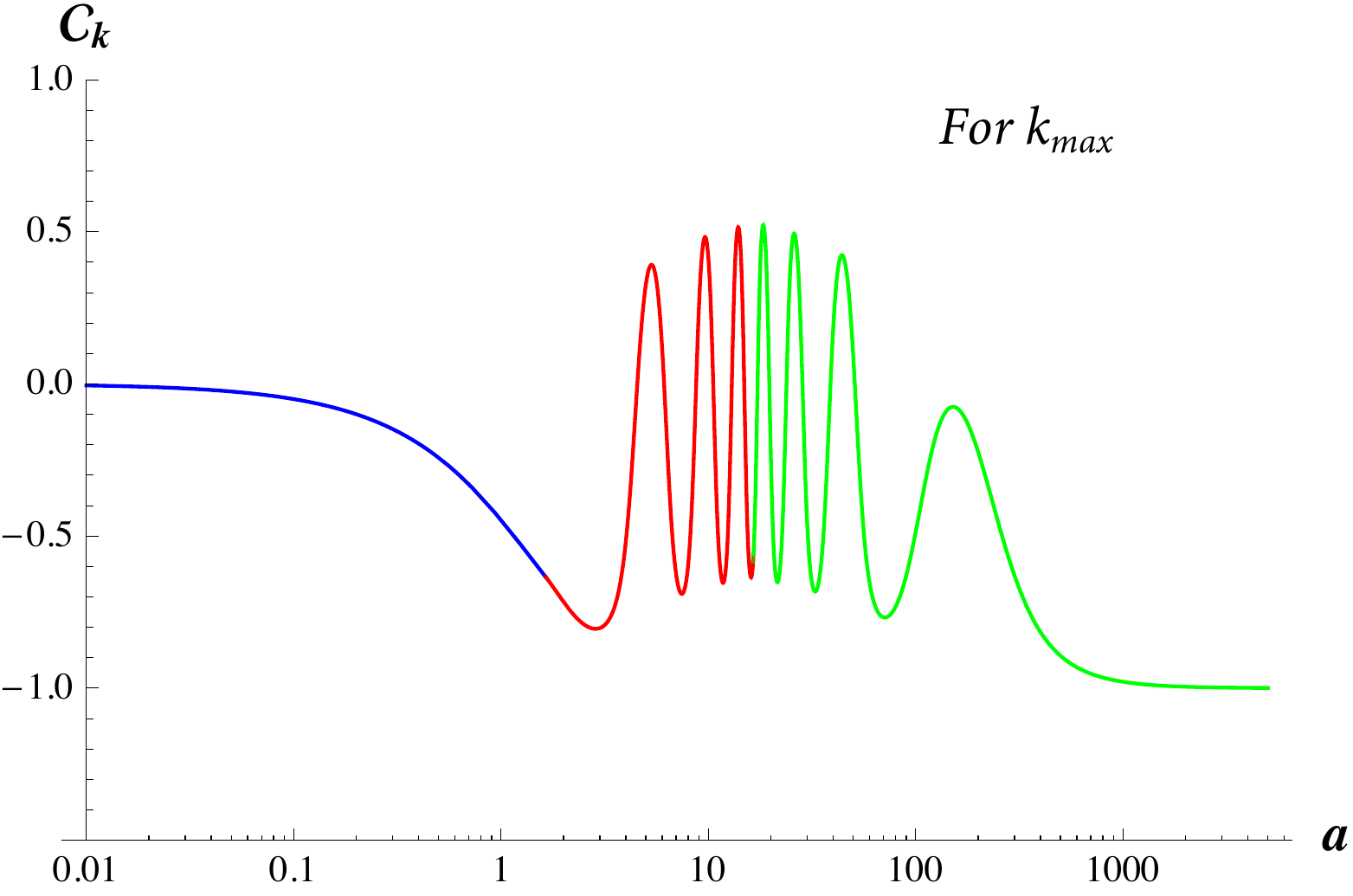}\hfill
\includegraphics[width=0.33\textwidth]{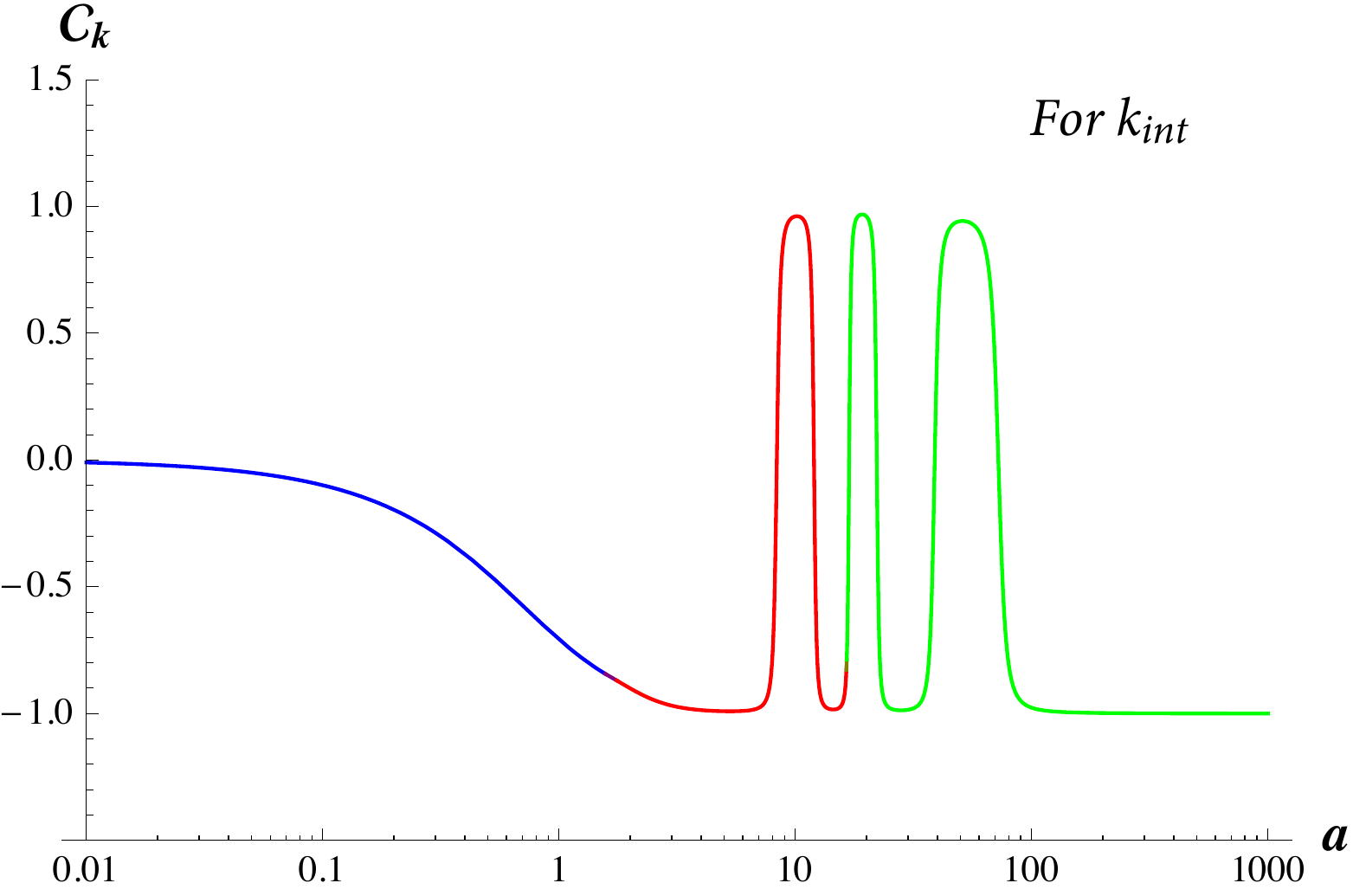}\hfill
\includegraphics[width=0.33\textwidth]{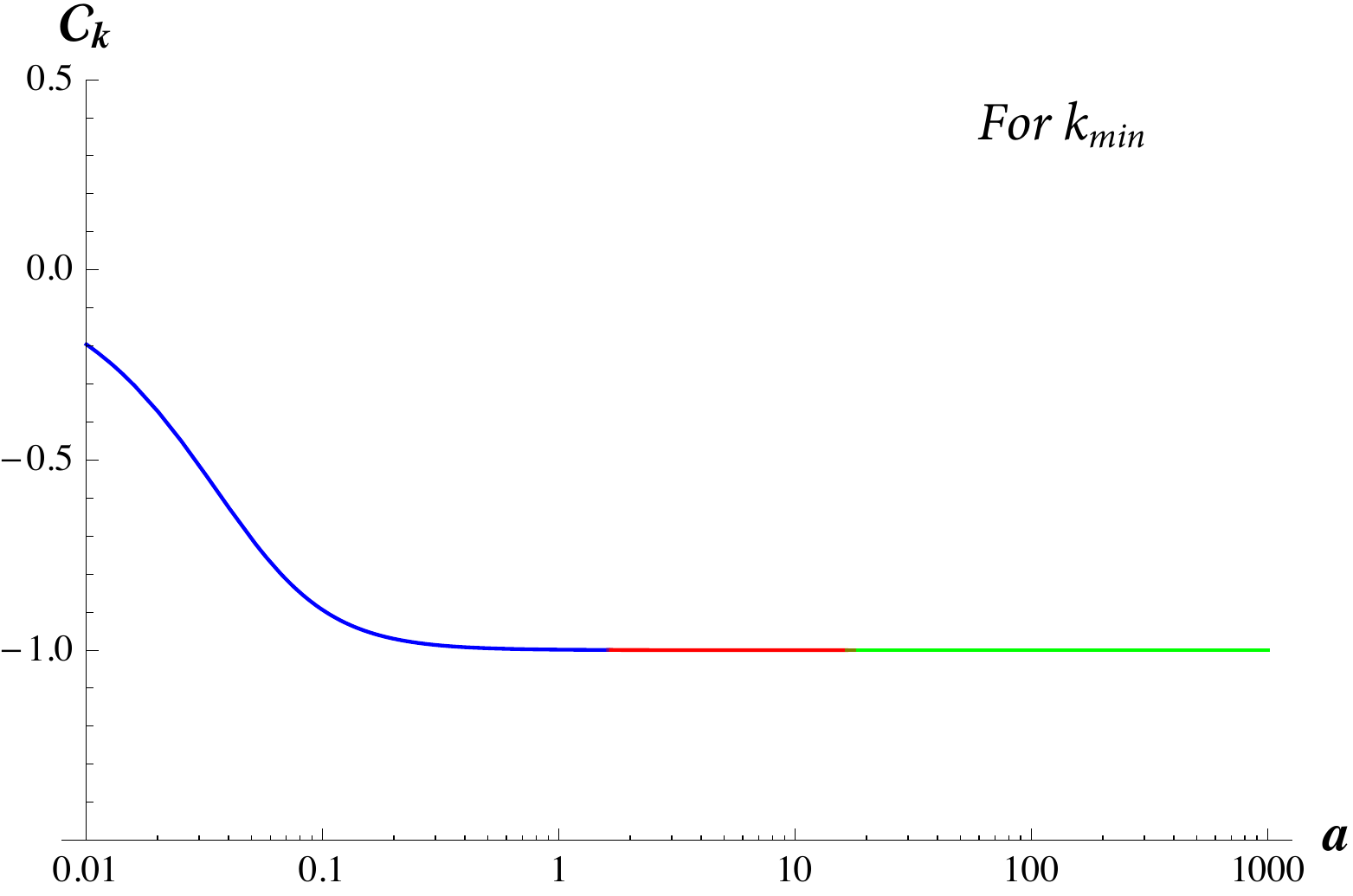}\hfill
\caption{Evolution  of the excitation parameter $z_{k}$, average particle number $\langle n_{k} \rangle$ and classicality parameter ${\cal C}_{k}$ with the scale factor for $\epsilon = 0.01$. The color scheme is as follows: blue $\rightarrow$ early inflationary phase, red $\rightarrow$ radiation dominated phase, green $\rightarrow$ late-time de Sitter phase. The modulus of the wave vector $k$ is fixed in each plot. Modes which have $k < k_{min}$ are always super-Hubble and those with $k > k_{max}$ are always sub-Hubble until they finally exit the late-time de Sitter phase. Any other intermediate mode ($k_{int}$) exit the Hubble radius in early de Sitter phase at $a_{ex_{1}}$ then enters in the radiation dominated phase at $a_{en}$ and finally re-exits in late-time de Sitter phase at $a_{ex_{2}}$. Here $k_{max} =1.64872 $, $k_{min} = 0.164872$, $k_{int} = 1$ for $z_k$ and ${\cal C}_k$ and $k_{int} = 0.5$ for $\langle n_{k} \rangle$ while $k = 10,~0.05 $ for lower and uppermost plots respectively, outside the band.}
\label{epsilon1}
\end{figure*}

\begin{figure*}[t!]
\includegraphics[width=0.40\textwidth,scale=0.2]{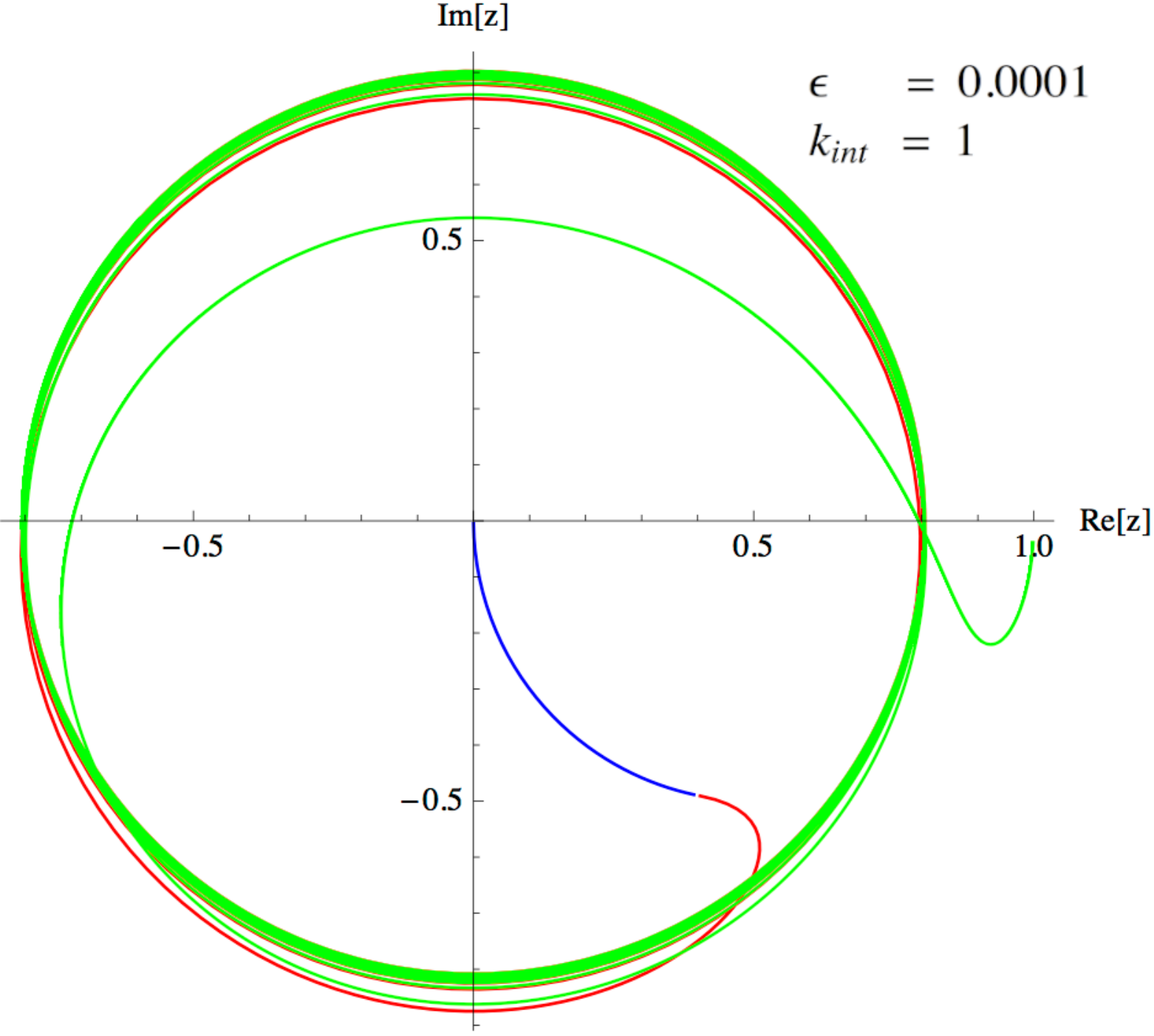}\hfill
\includegraphics[width=0.49\textwidth]{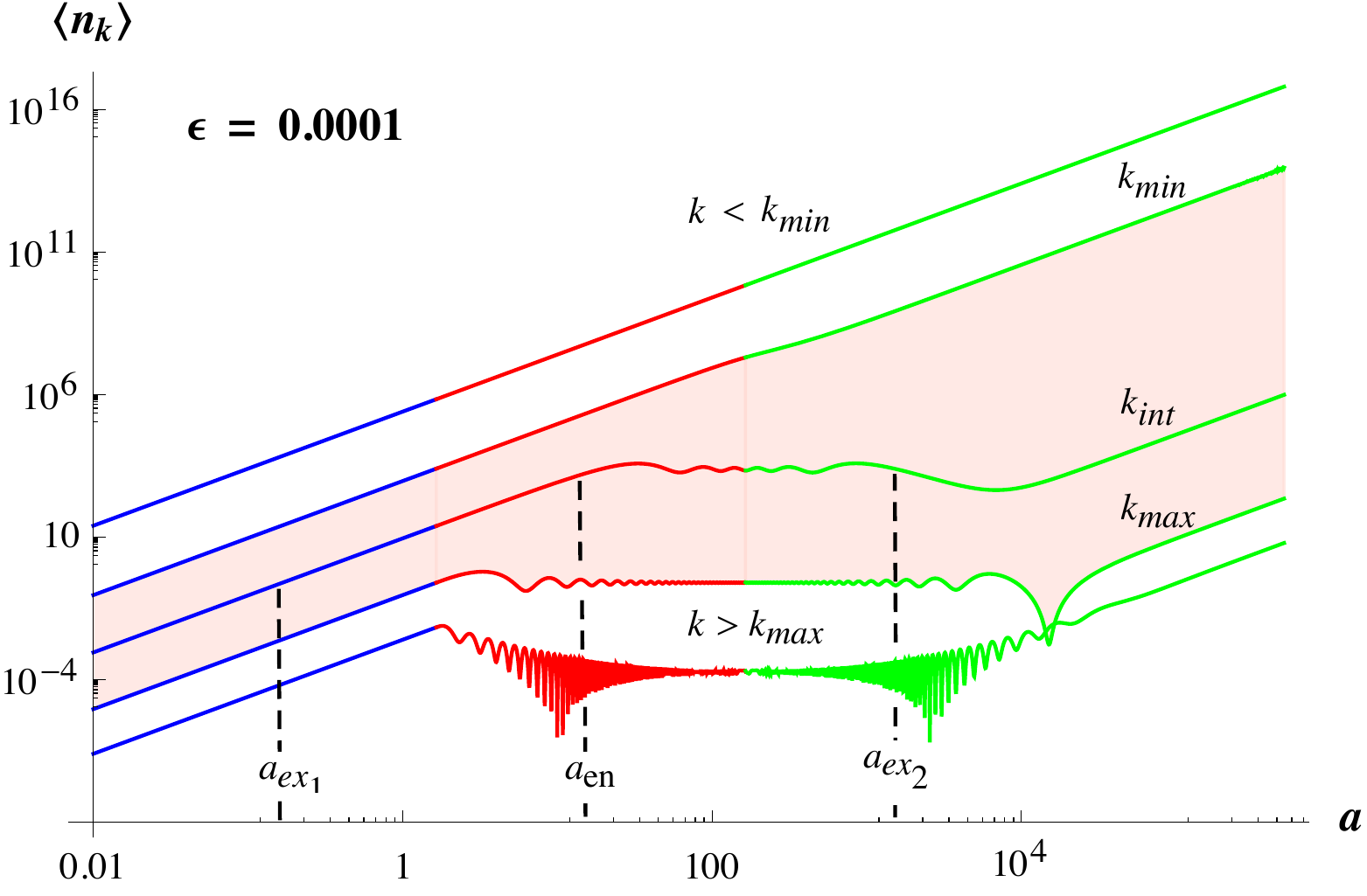}
\includegraphics[width=0.33\textwidth]{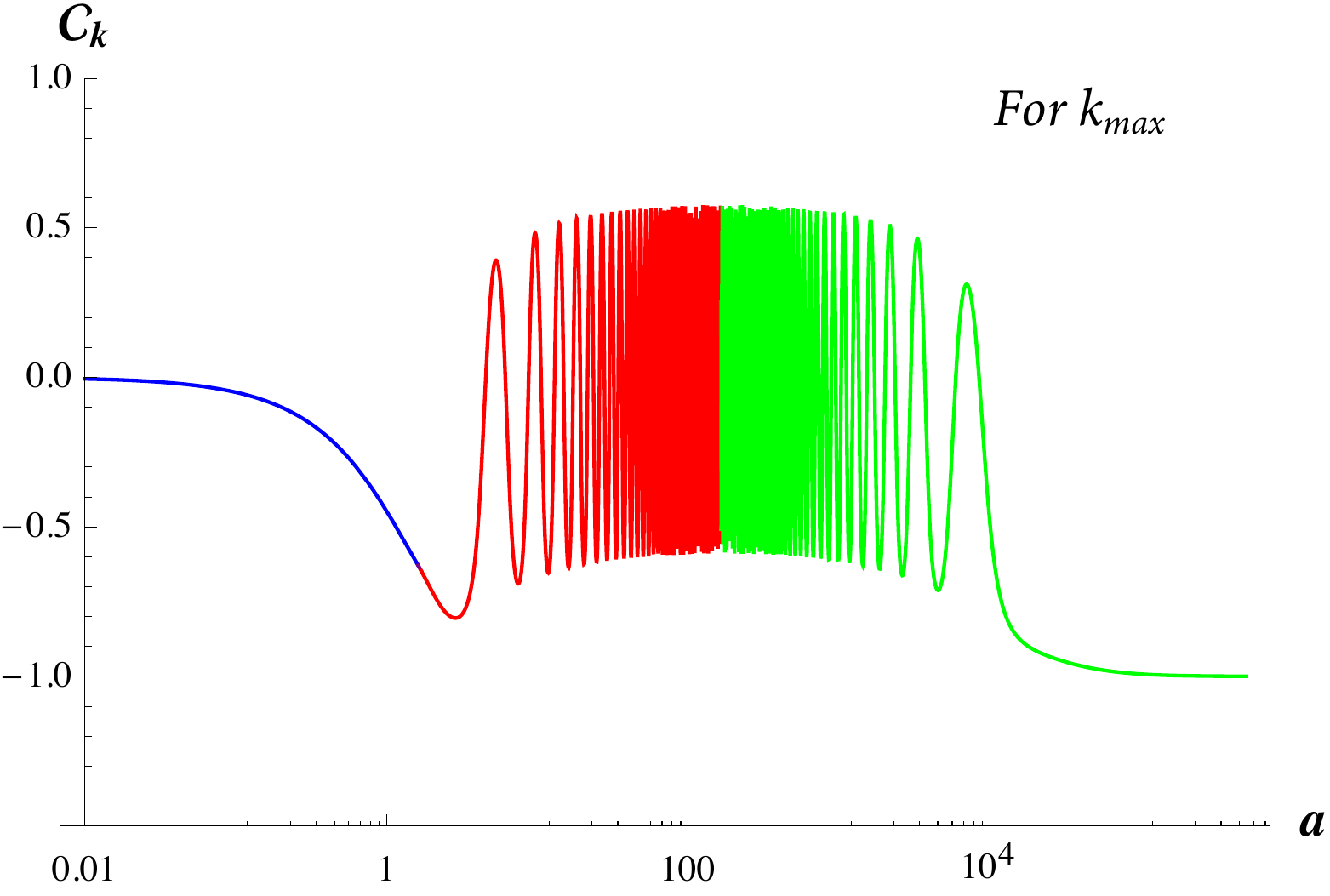}\hfill
\includegraphics[width=0.33\textwidth]{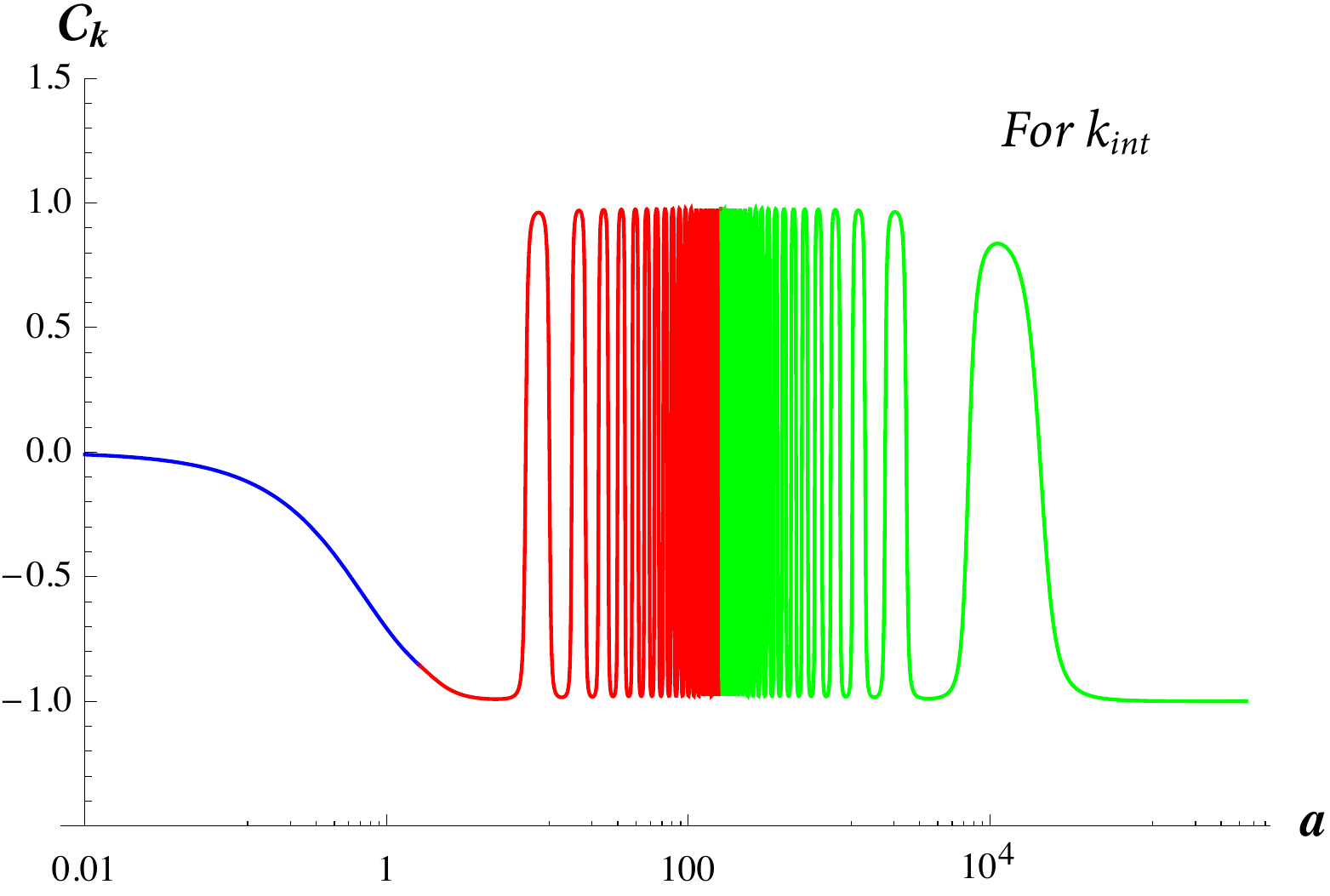}\hfill
\includegraphics[width=0.33\textwidth]{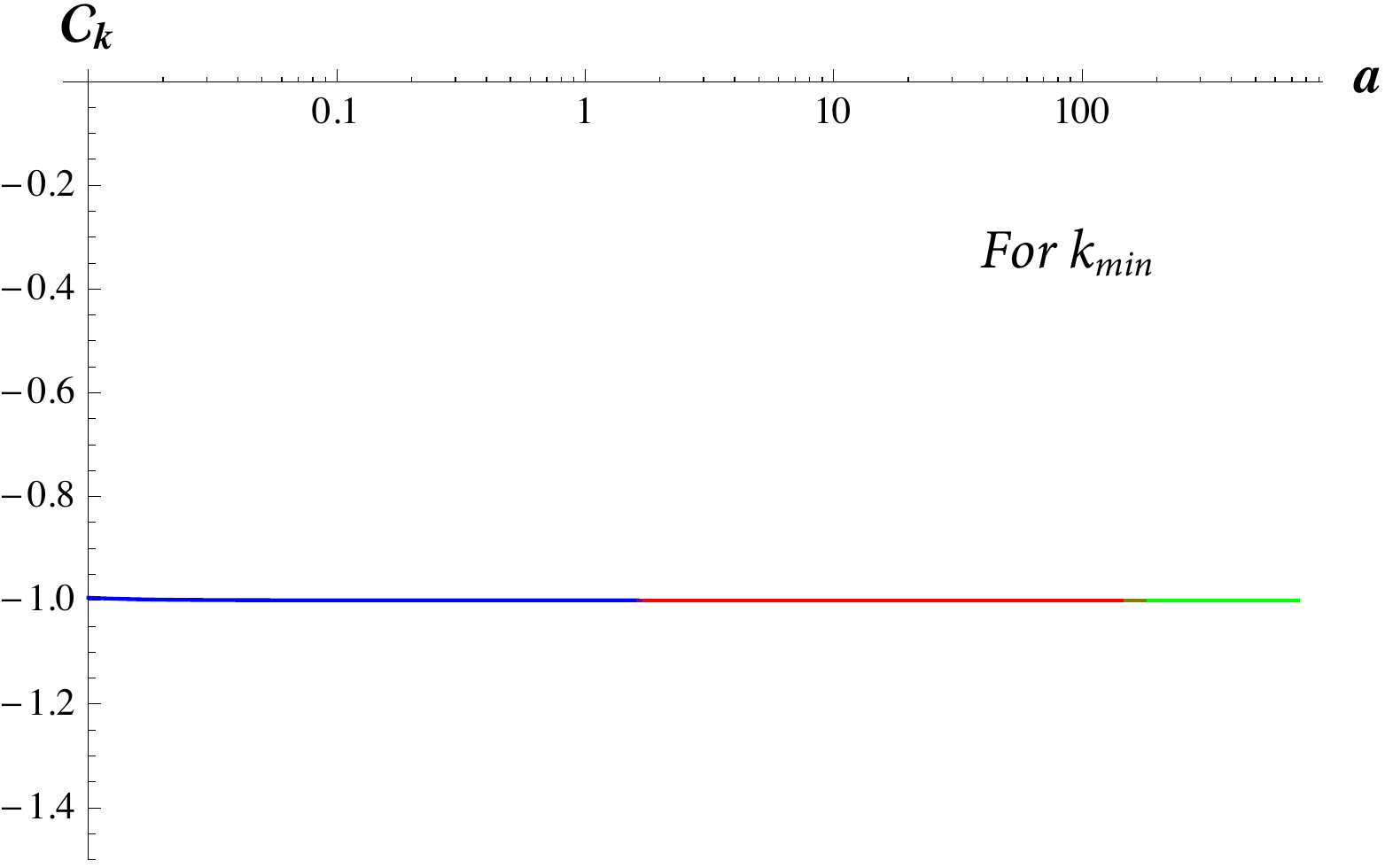}
\caption{Evolution of the excitation parameter $z_{k}$, average particle number $\langle n_{k} \rangle$ and classicality parameter ${\cal C}_{k}$ with the scale factor for $\epsilon = 0.0001$. Color scheme and notations are same as \fig{epsilon1} but in this case $k_{max} = 1.6487$, $k_{min} = 0.01648$ and $k_{int} = 0.167$. Outside the band, $k = 10.0$ (lower) and $k = 0.001$ (upper). For $z_{k}$ and $\mathcal{C}_{k}$ we have taken $k_{int} = 1.0$.}  
\label{epsilon2}
\end{figure*}

The excitation parameter $z_{k}$ is calculated separately for three stages from \eq{zmu}. Since we have three different expressions for $\mu_{k}$, given by \eq{muk1}, \eq{muk2} and \eq{muk3} in different stages there exist three corresponding results for $z_{k}$ viz., $z^{(1)}_{k},~z^{(2)}_{k}$ and $z^{(3)}_{k}$ which by construction match at the transition points. Then it is straight forward to calculate $\langle n_k \rangle$ and ${\cal C}_{k}$ separately for each stage and plot them collectively as in \fig{epsilon1} and \fig{epsilon2}.  

\fig{epsilon1} depicts the case with $\epsilon=0.01$. The excitation parameter starts from $z_{k}^{(1)}=0$ (shown in blue) in the inflationary phase and it gradually shifts toward $z_{k}^{(3)}=1$ (shown in green) in the late-time de Sitter. As we have already mentioned, the non-zero values of the excitation parameter ($z_{k}$) implies that the system has departed from its initial adiabatic vacuum state given by $z_{k}^{(1)}=0$. The appearance of the radiation dominated phase in between makes $z_{k}^{(2)}$ (shown in red) to follow a circular trajectory for a while and --- somewhat interestingly --- even after this phase has ended, $z_{k}^{(3)}$ (green) continues to exhibit the same behavior, before it changes its direction to reach unity. This behavior of $z_k$ is particularly interesting because for a single de Sitter case $z_{k}$ does not follow a circular path \cite{gaurang2007} which is clearly a `residual memory' of the radiation dominated phase. That is, despite the fact that the background spacetime has already made a transition from radiation to late-time de Sitter phase, the scalar field does not `know' about this for a while until $z_{k}^{(3)}$ leaves the circular trajectory. This is a  signature of non-adiabatic behavior that we mentioned above.

The above characteristics of $z_{k}$ have nontrivial consequences on $\langle n_{k} \rangle$ and ${\cal C}_{k}$. It is known that $\langle n_{k} \rangle$ follows a power law (and hence appears as a straight line in logarithmic plot) for pure de Sitter phase as previously found in \cite{gaurang2007}. This is because the average number of particles that are being created in this phase increases monotonously with the expansion until it reaches a mode dependent maximum value at the point of transition. This holds for $a \le e^{1/2}$ since the initial de Sitter phase (blue) has no prior information about the future transition. After the first transition, once the universe is in the radiation dominated phase (red), the behavior of $\langle n_{k} \rangle$ changes dramatically. This change basically depends upon the modulus of the wave vector ($k$). For super-Hubble modes ($k < k_{min}$), the power-law behaviour of $\langle n_{k} \rangle$ is unaffected because these modes have exited the Hubble radius in the inflationary phase itself. The $\langle n_{k} \rangle$ of  modes corresponding to $k > k_{min}$ starts {\it oscillating} once the universe changes to radiation dominated phase. This amplitude of the oscillation increases with $k$ and is most pronounced for sub-Hubble modes. One can directly relate this oscillatory behavior of $\langle n_{k} \rangle$ with the circular trajectories of $z_{k}$. Just as in the case of $z_k$, for $\langle n_{k} \rangle$ also, we see the effect of radiation dominated phase is `remembered' in the late-time de Sitter phase (green); in fact, this effect persists for quite some time. Note that in a pure de Sitter universe $\langle n_{k} \rangle$ has no oscillations. Therefore appearance of oscillations in the late-time de Sitter phase is a residual effect of the radiation dominated era. However, for large $a$, after the relevant modes become super-Hubble, this oscillatory behavior in $\langle n_{k} \rangle$ settles down to power-law behaviour. One important observation that follows from \fig{epsilon1} and \fig{epsilon2} is that irrespective of the value of $\epsilon$ the oscillation in $\langle n_{k} \rangle$ approaches a saturation value toward the end of the radiation phase. In fact, in the absence of the late-time de Sitter phase, $\langle n_{k} \rangle$ remains to be fixed at this saturated value forever (as it happens for a pure radiation-dominated universe \cite{gaurang2007}). The latter phase makes $\langle n_{k} \rangle$ to shift from its saturated value it would have reached in the radiation dominated phase. However, as we pointed out before, because of the `memory' it is not possible to instantaneously drive the system away from this saturation value. Some amount of time has to be spent in the late-time de Sitter phase for this to occur, which varies from mode to mode and turns out to be larger for sub-Hubble ($k > k_{max}$) regime. 

Interestingly, this saturation in average particle number has a similarity with the particle creation in electric field that was encountered earlier in \cite{gaurang2007, gaurang2008}. In the presence of constant electric field, at late times, $\langle n_{k} \rangle$ becomes nearly constant \cite{gaurang2007}. For time dependent electric field there are two cases which were discussed in \cite{gaurang2008}:  For smaller values of a dimensionless parameter $\sqrt{qE}t$ (where $q$ is the charge and $E$ is the electric field), the asymptotic mean particle number depends upon the duration for which the field ($E$) was nonzero, whereas,  for considerably larger values of $\sqrt{qE}t$, the final particle content becomes independent of this duration. Similar results hold in our case. The asymptotic average particle number near the end of radiation phase fluctuates for smaller lifetime of the radiation phase (for example, with $\epsilon=0.01$ in  \fig{epsilon1}) and for relatively larger lifetime it becomes constant (like in  \fig{epsilon2} with $\epsilon = 0.0001$). We shall return to this discussion once again in the next subsection where we deal with analytical results.     

The classicality parameter ${\cal C}_k$ starts from zero in the beginning of inflationary phase and depending upon $k$ and $\epsilon$ it shows different characteristics as depicted in \fig{epsilon1} and \fig{epsilon2}. Any mode with $k = k_{int}$ which lies within the $[k_{min},\,k_{max}]$ band, tends to a classical description near the end of the inflationary phase as ${\cal C}_k \rightarrow -1$ but as the universe makes a transition to radiation phase, it starts oscillating. These oscillations last  during the radiation phase as well as in the beginning of the late-time de Sitter phase. In the late-time de Sitter phase when a mode exits the Hubble radius at large $a$ one finds  ${\cal C}_k$ saturates at -1. This property of the mode establishes a connection between the classicality and its Hubble exit. After the Hubble  exit, in the early and late-time de Sitter  ${\cal C}_k \rightarrow -1$ and modes behave classically (as expected). But in between, when the mode is sub-Hubble, it oscillates and remains away from classical description. To clearly understand this aspect we have plotted ${\cal C}_{k}$ by considering two modes which are sub-Hubble  (for the first two phases) and is super-Hubble  (once it exits from the inflationary phase). The sub-Hubble mode does not reach -1 in the first two phases but once it exits the Hubble radius in the late-time de Sitter phase it reaches that value. On the other hand, the super-Hubble mode, once it exits from the initial de Sitter phase remains always super-Hubble and saturates with ${\cal C}_k \rightarrow -1$. This relation of classicality and Hubble exit is, of course, known (\cite{brand}-\cite{othr} and \cite{lyth2008}) in the context of primordial perturbations but our procedure provides a quantitative measure of degree of classicality. 

To understand the oscillatory nature in the average particle number in \fig{epsilon1} and \fig{epsilon2} one should again refer to the behavior of classicality parameter. The fluctuations in ${\cal C}_k$ imply that the system is in the quantum domain and the notion of average number of particles at a particular instant is not well defined in the `classical' sense. At most one can ask, in loose sense, a time averaged value of  $\langle n_{k} \rangle$ and interpret this as the number of particles produced during that time interval. The particle content is well-defined and has the intuitive behaviour of monotonic increase only when degree of classicality is high which is precisely what we would expect. While our formalism allows us to define $\langle n_{k} \rangle$ at any time, one cannot really think of them as `particles' when its value is oscillatory. This is precisely what happens  when degree of classicality is low (as is to be expected) and particle definition becomes ambiguous. As we stressed right at the beginning of the paper we do not want to over emphasize any given notion of particle in a strong field regime; in stead we want to correlate the behaviour of a well-defined parameter (measured by $\langle n_{k} \rangle$) with degree of classicality. this study confirms our intuitive expectations and adds strength to the interpretation of both our definition of particle content and degree of classicality. 

Almost all physical features of \fig{epsilon1} remain intact for $\epsilon = 0.0001$ as is evident from \fig{epsilon2}. The only difference arises in terms of number of rotations encountered in the plot of $z_{k}$. Since for this case the radiation phase has a comparatively longer lifetime the oscillations of $\langle n_{k} \rangle$ and ${\cal C}_k$ persist for a longer duration. In fact, as we make $\epsilon$ smaller, this winding number increases substantially and becomes unmanageably high for cosmological value of $\epsilon \sim 10^{-54}$. We shall now look at this case analytically, taking suitable limits, in the next subsection.

\subsection{Analytic limits in different regimes}
\label{subsec:analyticlimits}
As we have already mentioned, the analytical results become increasingly complex as we proceed with the evolution of the universe and thus require suitable approximations for the final late-time de Sitter phase. For the inflationary phase, the expressions for various quantities are simple, and are given by
\be
\label{z1}
z_k^{(1)} = \frac{a H}{a H+2 i k}
\ee

\bes
\label{n1}
  \langle n_{k}^{(1)} \rangle &=& \frac{a^2 H^2}{4 k^2} \\
  {\cal C}_{k}^{(1)} &=& -\frac{a H}{k \sqrt{\frac{a^2 H^2}{k^2}+1}}
\ees
\\
These match with the results found earlier in \cite{gaurang2007}. For any value of $k/H$, in the logarithmic plot, $\langle n_{k}^{(1)} \rangle\propto a^2$ is a straight line  as shown in \fig{epsilon1} and \fig{epsilon2} (blue lines). Also, as expected, none of the expressions above depend upon $\epsilon$. From the expression of the classicality parameter it is obvious that super-Hubble modes with $aH/k >> 1$ have ${\cal C}_{k}^{(1)} \approx -1$ and behave classically. However, all other modes remain away from classical description. In this phase, we can also calculate the \textit{comoving} energy density by multiplying \eq{n1} by $k$ (since $\omega_k = k$) and integrating over $\mathrm{d}^3k$ as
\bes
\mathcal{E}^{(1)} (a) &=& \f{a^2H^2}{8\pi^2  }\int_{k_{a}}^{k_{b}} k \dd k\nn\\
&=& \f{a^2H^2}{16\pi^2}(k_{b}^2 - k_{a}^2).
\ees 
The above expression has a UV divergence which is usual in the case in quantum field theory and requires a cutoff. The origin and cure for divergences in de Sitter phase is still a matter of debate in the literature but for our purpose we shall just introduce a cut-off. We can choose $k_a = k_{min}$ and $k_b = k_{max}$ for the energy in the band $[k_{min}$,\,$k_{max}]$. For the cosmological case, this band is wide enough to be interesting with $k_{min} = H(e\epsilon)^{1/2}$ and $k_{max}= H e^{1/2}$. The background comoving energy density is $\mathcal{E}^H_{bg} = 3H^2/8\pi G$. So the ratio of the energy densities at the end of inflation is,
\be
\label{comp1}
\f{\mathcal{E}^{(1)} (a_r)}{\mathcal{E}^{H}_{bg}} = \f{e^2H^2 G}{6\pi}(1 -\epsilon) \sim L_p^2 H^2 
\ee
where $L_p$ is the planck length and we have $\epsilon<<1$. With $E_p = 10^{19}$ GeV and $E_{GUT} = 10^{15}$ GEV, this ratio is of order $10^{-16}$ which shows that $\mathcal{E}_{\mathrm{inf}}<<\mathcal{E}_B$. As a result  the particle creation in the band we are studying does not substantially backreact on the background geometry. 

In the radiation dominated phase, the exact analytical results for $z_k^{(2)}$ and $\langle n_{k}^{(2)} \rangle$ are a little cumbersome and are given by
\bwt
\bes
\label{z2}
z_k^{(2)} &=& \frac{e H \left(e^{\frac{2 i k}{\sqrt{e} H}} H (e H+2 i a k)-e^{\frac{2 i a k}{e H}} \left(e H^2+2 i \sqrt{e} H k-2 k^2\right)\right)}{-e^{2+\frac{2 i k}{\sqrt{e} H}} H^3+e^{\frac{2 i a k}{e H}} (e H-2 i a k) \left(e H^2+2 i \sqrt{e} H k-2 k^2\right)} \\ \label{n2}
  \langle n_{k}^{(2)} \rangle &=& \frac{e^2 H^2}{8 a^2 k^6} \left(e^2 H^4+2 \left(a^2 H^2 k^2+k^4\right) -\left(e^2 H^4+4 a \sqrt{e} H^2 k^2-2 e H^2 k^2\right) \text{Cos}\left[\frac{2 \left(-a+\sqrt{e}\right) k}{e H}\right] \right. \notag \\
 && \left.  -2 kH \left(\left(-a+\sqrt{e}\right) e H^2+2 a k^2\right) \text{Sin}\left[\frac{2 \left(-a+\sqrt{e}\right) k}{e H}\right]\right). \ees
\ewt 
Note that although the scale factor in a radiation dominated phase has an explicit $\epsilon$ dependence in \eq{aofeta}, both $z_{k}^{(2)}$ and $\langle n_{k}^{(2)} \rangle$ are not explicit functions of $\epsilon$. Let us now examine the behavior in the special cases which are of physical interest. For super-Hubble modes ($k < k_{min} =  (\epsilon e)^{1/2}$), it turns out that 
\be
\langle n_{k}^{(2)}\rangle  \approx \frac{a^2 H^2}{4 k^2} + \mathcal{O}(k^2)
\ee
which exactly matches with $\langle n_{k}^{(1)} \rangle$ in \eq{n1}. Not surprisingly, for these modes the log plot is a straight line with slope 2 as shown in \fig{epsilon1} and \fig{epsilon2}. One can always change $\epsilon$ to compare with the cosmological case and in such a case $k_{min}$ is also shifted to a much lower value which is proportional to $\epsilon^{1/2}$. On the other hand for sub-Hubble modes ($k > k_{max}$) the argument of the Sine and Cosine terms in \eq{n2} get bigger and the oscillatory nature takes over from monotonic behaviour. Asymptotically, as the universe approaches the end of radiation phase  (for $a \rightarrow e^{1/2}/\epsilon^{1/2}$), the $\langle n_{k}^{(2)} \rangle$ tends to a saturation value
\be 
\label{n2satu}
{\langle n_{k}^{(2)} \rangle}_{sat} \approx \frac{e^2 H^4}{4 k^4}
\ee
when we ignore the smaller oscillatory terms. This is approximately the average number of particles at the end of radiation phase and correspond to the saturated regime of the plots in \fig{epsilon1} and \fig{epsilon2}. Again, it is straightforward to find the average energy density due to these particles by multiplying (\ref{n2satu}) by $k^3dk/2\pi^2$ and integrating over all $k$ (from some $k_a$ to $k_b$) to give:
\be
\label{ensatu}
{\cal E}^{(2)}_{sat} = \frac{e^2H^4}{8\pi^2}\ln\Big|\frac{k_b}{k_a}\Big|.
\ee
The average energy density per logarithmic mode is a constant and is proportional to the fourth power of the inflationary Hubble parameter. We can again compare its value with the comoving background energy density at the end of radiation phase which is
\be
\mathcal{E}^{\mathrm{rad}}_{bg} = \f{3H^2 e^2}{8\pi L_p^2}
\ee
to get
\be
\f{{\cal E}^{(2)}_{sat}}{\mathcal{E}^{\mathrm{rad}}_{bg} } = \f{L_p^2 H^2}{3\pi} \ln(\epsilon^{-1/2}) \sim L_p^2 H^2
\ee
which essentially remains almost the same as in \eq{comp1} since there is not much particle creation during the radiation phase due to saturation. The backreaction due to the modes in the band we are studying is not a concern for the evolution of background geometry. 

For the late-time de Sitter phase, we have given the exact expression for $z_k^{(3)}$ in the ~\app{app:z3}. (The exact analytic expression for ${\langle n_{k}^{(3)} \rangle}$ is too cumbersome to offer any insight and hence is not included.) Let us consider this expression in the appropriate limits. First, note that due to matching conditions ${\langle n_{k}^{(3)} \rangle}$ is equal to the value of ${\langle n_{k}^{(2)} \rangle}$ as given in \eq{n2satu} at the beginning of this phase for all $k$. Further, for any $k$, at late-times ${\langle n_{k}^{(3)} \rangle}$ varies as $a^2$ leading to straight lines of slope 2 in log-log plots. Again, for all super-Hubble modes with $k<k_{min} = (\epsilon e)^{1/2}$ in the small $k$ limit, this  behavior continues with
\be
z_k^{(3)} \approx 1-\frac{2 k^2}{a^2 H^2}-\frac{2 i k}{a H \epsilon }
\ee
which then gives
\be
\langle n_{k}^{(3)}\rangle  \approx \frac{a^2 H^2}{4 k^2} 
\ee
with the functional form of ${\langle n_{k}^{(3)} \rangle}$ matching ${\langle n_{k}^{(1)} \rangle}$given in \eq{n1}. These features are remain valid for  the realistic cosmological value of $\epsilon$, for which the numerical value of $k_{min} \sim \epsilon^{1/2} \sim 10^{-27}$ is extremely small. 

The situation is more complicated for the field modes with $k>k_{min}$. Some of these modes fall in the intermediate band $k_{min}<k<k_{max}$ and others remain sub-Hubble until they exit late-time de Sitter. To quantify the modes in the intermediate band is difficult analytically, but we can study the other limit of $k>k_{max}$. For these modes, with small $\epsilon$  and  large $a$ (since $a > (e/\epsilon)^{1/2}$) for the third region, the dominant terms of ${\langle n_{k}^{(3)} \rangle}$  are given by

\begin{figure}[t!]
\includegraphics[scale=0.5]{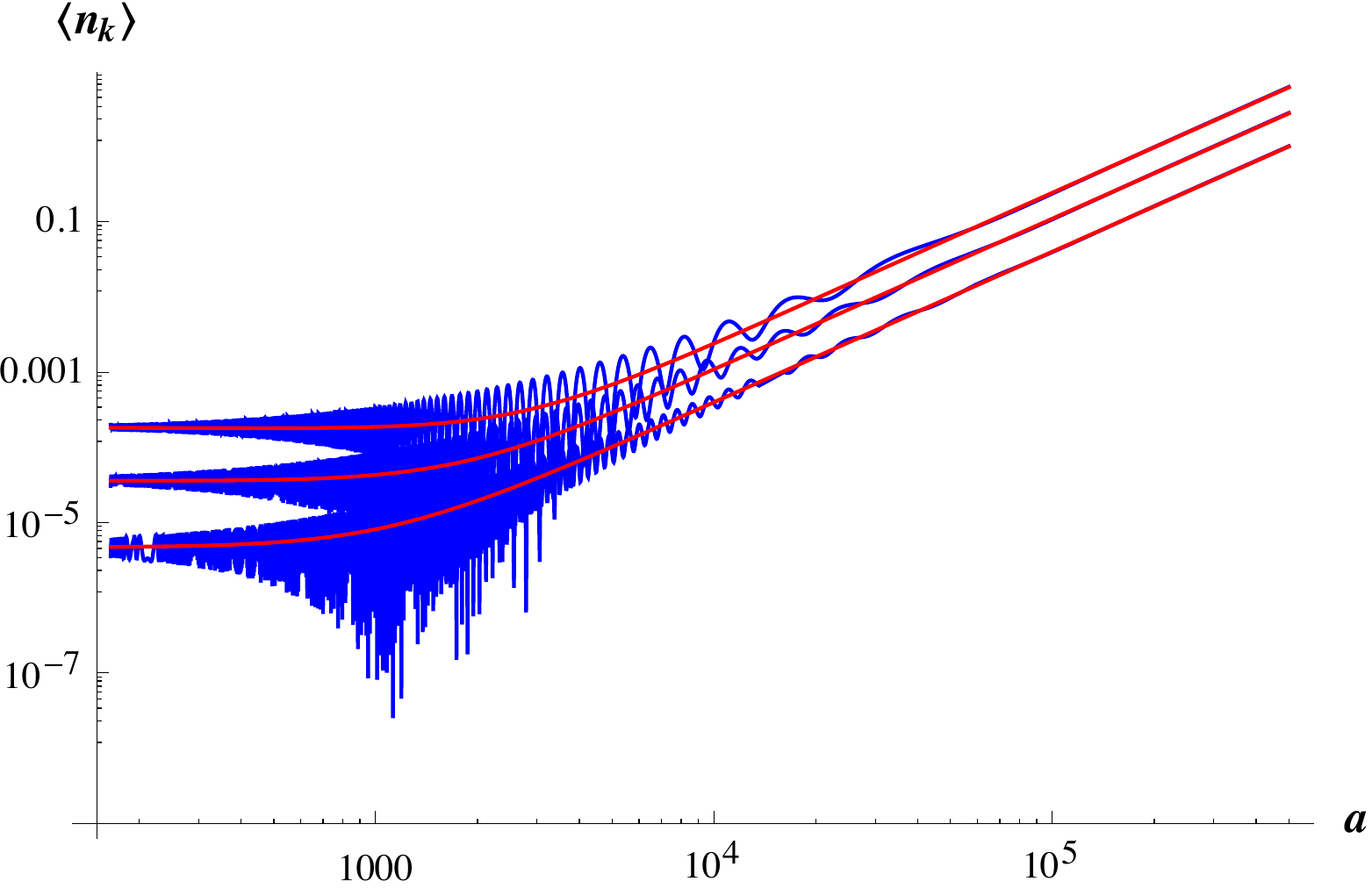}
\caption{Comparison of the average particle number $\langle n_{k}^{(3)}\rangle$ (shown in blue) with its average value ignoring the oscillations (shown in red) as given by the expression in~\eq{analyticn3} in the large $k$ limit. Here $\epsilon = 0.0001$ and $k/H = 10,15, 25$ increasing downwards.}
\label{analyticcompare}
\end{figure} 

\bes
\label{analyticn3}
{\langle n_{k}^{(3)} \rangle} &\approx& \frac{e^2 H^4}{4 k^4} - \frac{2H^4 e^{3/2} a\epsilon}{4 k^4}\nn\\
&&+\frac{H^2  \left(e^2H^4-e H^2k^2+k^4\right)a^2\epsilon^2}{4 k^6}
\ees
This  expression summarizes the behavior of particle content the third region when the oscillations are averaged out in and is shown in \fig{analyticcompare}. The zeroth order term is the saturation value of ${\langle n_{k}^{(2)} \rangle}$ which ${\langle n_{k}^{(3)} \rangle}$ picks up for relatively small $a$. Later on, when when $aH\epsilon/k >> 1$, the $a^2$ behaviour becomes the dominant feature; the behaviour shown in~\fig{epsilon1} and~\fig{epsilon2} corroborates this, i.e., when $a > k/H\epsilon$, the oscillations die down and monotonic behaviour arises. 

\begin{figure}[t!]
\includegraphics[scale=0.6]{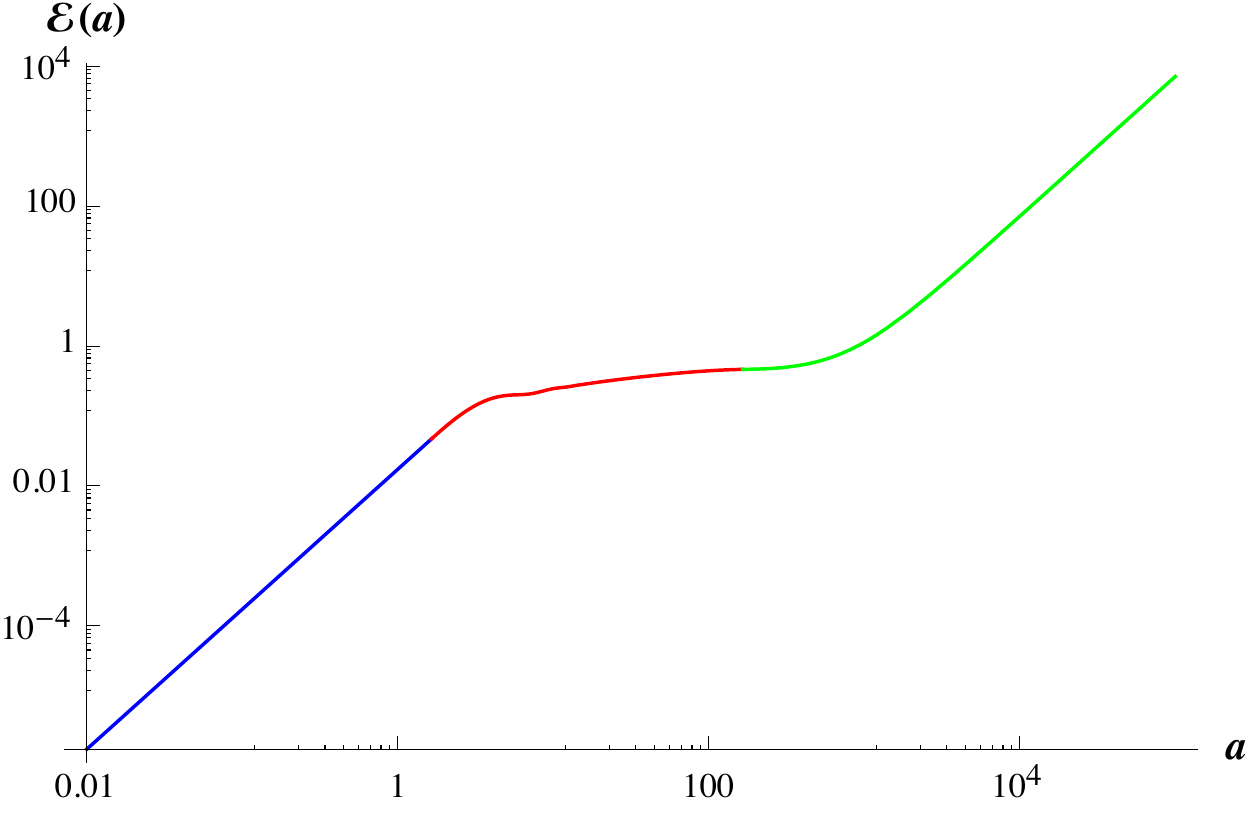}
\caption{Evolution of the energy density of the field in the $[k_{min}$,\,$k_{max}]$ band as given by \eq{energy} for $\epsilon = 0.0001$ and $H=1$. The color scheme is the same as in previous figures.}
\label{energydensity}
\end{figure}

Finally, we plot the energy density of the field in the $[k_{min},\,k_{max}]$ band given by
\be
\label{energy}
\mathcal{E} = \f{1}{2\pi^2}\int_{k_{min}}^{k_{max}} \langle n_k\rangle k^3 \dd k
\ee
in \fig{energydensity} for $\epsilon = 0.0001$ and $H=1$. The energy density increases sharply (red) in the inflationary phase due to significant  particle creation in this phase and is then almost constant in the radiation phase (blue) when  particle creation is insignificant. The saturated remnant is also seen early in the third phase (green). The energy density of the field is quite low as compared with the comoving background energy density and does not pose any backreaction issues for the modes we have studied.   

\subsection{Implications for generation of perturbations from inflation}
\label{subsec:implications}

Finally we want to comment briefly on the connection between this work and the generation of perturbations in the inflationary scenario. In the standard approach to this problem, one computes the quantum fluctuations of the field at the time when the mode exits the Hubble radius in the initial de Sitter phase. This is usually done in the form of the two-point function $\langle 0|\phi(x)\phi(y)|0\rangle$. The Fourier transform of this  C-number is then identified with the stochastic fluctuations of a \textit{classical} random field at the time of re-entry of the mode to the Hubble radius (see e.g., p 637 of \cite{gravitationTP}). The usual justification for this procedure is based on two factors: (a) The modes behave classically once they are well outside the Hubble radius. (b) It is assumed that once they become classical they stay classical and hence can be described by standard perturbation theory after they re-enter the Hubble radius. 

Though this scenario is by now widely accepted (and the results of such a computation agrees well with observations), it must be noted that  the quantum to classical transition of the density perturbations is still not completely well understood from a conceptual point of view. 
We believe the approach introduced this paper will throw more light on this issue. 
The key point is that we now have an intuitively clear, quantitative, measure for the classicality of the fluctuations and we need not deal with a `two-level' description of the fluctuations being either fully classical or fully quantum mechanical. We will confine ourselves to brief comments here and hope to address this issue more comprehensively in a separate work. 

Our discussion of the field modes \textit{does} confirm the standard assumption (a) above, viz. that modes become classical when they leave the Hubble radius. What is more is that we can quantify the degree of classicality as the universe evolves. But our analysis also shows that, when the mode re-enters the Hubble radius the the degree of classicality does not stay constant but rapidly oscillates. This fact can have important implications for structure formation scenarios which are based on assumption (b) above, viz. that once the fluctuations are classical, they remain classical. But we must stress that, in the work reported here, we have treated the quantum field as purely a test field and did not incorporate the back reaction due to perturbations in the geometry. In the correct approach we need to take into account both the metric fluctuations (in particular the scalar mode representing the Newtonian potential in a particular gauge) and the field fluctuations as a coupled system (see e.g, p. 636 of \cite{gravitationTP}). It seems natural to treat the scalar mode perturbations of geometry as classical, in which case, we have the standard scenario of a quantum system interacting with a classical one and we need to consider issues like decoherence (which is often invoked to explain the classicality of perturbations though this may not be completely satisfactory). Further, we also note that the classicality parameter oscillates rapidly when the modes have re-entered the Hubble radius which is a different behaviour compared to the one exhibited when they were originally sub-Hubble radius in the inflationary phase. (This is clearly seen in the behaviour of $z_k$ in all the figures). This difference tells us that the two situations are not identical and our interpretation needs to take this difference into account. So the question cannot be answered only based on the results of the work described here and we hope to extend it, taking into consideration the back reaction of geometry both in the quantum and classical regimes as well as the oscillations of the classicality parameter.

Similar comments apply to the study of the tensor components of the metric perturbations because they obey essentially the same type of wave equation as a scalar field in the cosmological context. Once again the conventional wisdom is that: (a) The gravity wave modes are quantum mechanical until they leave the Hubble radius in the inflationary phase but become classical when they are outside the Hubble radius.
(b) Once they become classical they remain classical as a stochastic gravity wave background today. The result in (a) can again be justified by an analysis similar to ours because the mathematics is essentially the same. But in tackling the issue of (b) it seems difficult to invoke effects due to back reaction [unlike in the case of scalar field modes] and it may be the oscillations of the classicality parameter which contains the relevant information. We hope to study these issues in detail in a future work. 

\section{Discussion and Conclusion}
\label{sec:dc}
We have provided a detailed analysis of {\it instantaneous} particle creation and quantum to classical transition for minimally coupled scalar field modes in the background of Friedman universe. We considered three main stages of expansion where the radiation dominated phase is sandwiched  between the early (inflationary) and late-time ($\Lambda$ dominated) de Sitter phase. Making the scale factor continuous and differentiable at the time of transition fixes the transition points in terms of Hubble parameters at two regimes. Following the general framework of \cite{gaurang2007} the field equations were solved in different regimes in Schr\"odinger picture. Boundary conditions were chosen to set the vacuum in a Bunch-Davies state in the asymptotic past of early de Sitter. The time evolution of these vacuum states is non-adiabatic and leads to particle creation.

The time dependent average particle number and classicality parameter for a given mode were expressed as a function of the scale factor. In the early de Sitter phase particle creation is uniform and $\langle n_k \rangle$ increases as $a^2$ for all modes. Super-Hubble modes ($k < k_{min}$), once they exit the Hubble radius, are  not affected by the expansion and $\langle n_k \rangle$ behaves the same for all $a$. In the radiation phase $\langle n_k \rangle$ oscillates for intermediate modes ($k_{min} < k <k_{max}$ which exit the Hubble radius in the inflationary phase  and enter in the radiation phase) as well as for the sub-Hubble ($k > k_{max}$)  modes. Interestingly,  even after the universe transits into the late-time de Sitter phase, oscillations persist as a {\it residual memory} of radiation dominated phase. For large $a$, once these modes become super-Hubble, $\langle n_k \rangle$ again shows power-law behavior. It was observed that $\langle n_k \rangle$ reaches a saturated value towards the end of radiation phase in such a way that the average energy of the created particles per logarithmic mode becomes constant. Interestingly, this property of saturation was found to be similar to the case of particle creation by time dependent electric fields in certain circumstances \cite{gaurang2008}. 

The classicality parameter (${\cal C}_k$) starts from zero (for vacuum state) in the early inflationary phase and with the expansion its magnitude approaches to unity  within the same phase. Once it becomes radiation dominated, ${\cal C}_k$  fluctuates rapidly between -1 and +1. This has a counterpart in the  oscillatory behavior of $\langle n_k \rangle$. The general picture that emerges from our study suggests that modes behaves classically when they are {\it super-Hubble} and for them the magnitude of ${\cal C}_k$ approaches to unity. This is also the regime in which the system is highly non-adiabatic and particle production is strongest. These features in ${\cal C}_k$ closely track the emergence of classicality at late times. 

\section*{Acknowledgements}
S.S. is supported by SPM grant from the Council of Scientific and Industrial Research (CSIR), India. T.P.'s research is partially supported by the J.C.Bose Research Grant of DST, India. The authors also acknowledge help from Kaustubh P. Vaghmare for help in Python coding.

\appendix
\section{Exact expression for $z_k^{(3)}(a)$}
\label{app:z3}
The exact expression for $z_k^{(3)}(a)$ is given by
\be
\label{z3exact}
z_k^{(3)}(a) = \mathcal{N}/\mathcal{D}
\ee
where the numerator and denominator are respectively,
\bwt
\bes
\mathcal{N} &=& H \left(-e^{\frac{4 i k}{\sqrt{e} H}+\frac{2 i k}{H \epsilon }+\frac{5}{2}} H^4 (a H \epsilon -2 i k) \epsilon ^{3/2}+e^{\frac{2 i \left(\epsilon +\sqrt{\epsilon }\right) k}{\sqrt{e} H \epsilon }+\frac{2 i k}{H \epsilon }+\frac{5}{2}} H^4 (a H \epsilon -2 i k) \epsilon ^{3/2}-a e^{\frac{2 i \left(\sqrt{\epsilon }+2\right) k}{\sqrt{e} H \sqrt{\epsilon }}+\frac{2 (a-1) i k}{a H \epsilon }+\frac{5}{2}} H^5 \epsilon ^{5/2}\right. \nn \\
&&\left. +a e^{\frac{2 i \left(2 \epsilon +\sqrt{\epsilon }\right) k}{\sqrt{e} H \epsilon }+\frac{2 (a-1) i k}{a H \epsilon }+\frac{5}{2}} H^5 \epsilon ^{5/2}+a e^{\frac{2 i \left(2 \epsilon +\sqrt{\epsilon }\right) k}{\sqrt{e} H \epsilon }+\frac{2 (a-1) i k}{a H \epsilon }+2} H^4 i k \epsilon ^2+2 a e^{\frac{2 i \left(\sqrt{\epsilon }+2\right) k}{\sqrt{e} H \sqrt{\epsilon }}+\frac{2 (a-1) i k}{a H \epsilon }+\frac{3}{2}} H^3 k^2 \left(\sqrt{\epsilon }-1\right) \epsilon ^2\right.\nn \\
&&\left. -i a e^{\frac{2 i \left(\sqrt{\epsilon }+2\right) k}{\sqrt{e} H \sqrt{\epsilon }}+\frac{2 (a-1) i k}{a H \epsilon }+2} H^4 k \left(2 \sqrt{\epsilon }-1\right) \epsilon ^2-4 i a e^{\frac{2 i k \left(\sqrt{e} (a-1)+a \left(\epsilon +2 \sqrt{\epsilon }\right)\right)}{a \sqrt{e} H \epsilon }} k^5 \epsilon -4 a e^{\frac{2 i \left(\sqrt{\epsilon }+2\right) k}{\sqrt{e} H \sqrt{\epsilon }}+\frac{2 (a-1) i k}{a H \epsilon }+\frac{1}{2}} H k^4 \epsilon \right.\nn \\
 &&\left. -2 i a e^{\frac{2 i \left(\sqrt{\epsilon }+2\right) k}{\sqrt{e} H \sqrt{\epsilon }}+\frac{2 (a-1) i k}{a H \epsilon }+1} H^2 k^3 (\epsilon -1) \epsilon -2 i e^{\frac{2 i \left(\epsilon +\sqrt{\epsilon }\right) k}{\sqrt{e} H \epsilon }+\frac{2 i k}{H \epsilon }+1} H k^3 (a H \epsilon -2 i k) \epsilon -3 i e^{\frac{4 i k}{\sqrt{e} H}+\frac{2 i k}{H \epsilon }+2} H^3 k (a H \epsilon -2 i k) \epsilon \right. \nn\\
&&\left. -2 e^{\frac{2 i \left(\epsilon +\sqrt{\epsilon }\right) k}{\sqrt{e} H \epsilon }+\frac{2 i k}{H \epsilon }+\frac{3}{2}} H^2 k^2 \left(\sqrt{\epsilon }+1\right) (a H \epsilon -2 i k) \epsilon +e^{\frac{2 i \left(\epsilon +\sqrt{\epsilon }\right) k}{\sqrt{e} H \epsilon }+\frac{2 i k}{H \epsilon }+2} H^3 k \left(2 \sqrt{\epsilon }+1\right) (2 k+a H i \epsilon ) \epsilon \right. \nn \\
&&\left. +4 e^{\frac{4 i k}{\sqrt{e} H}+\frac{2 i k}{H \epsilon }+\frac{3}{2}} H^2 k^2 (a H \epsilon -2 i k) \sqrt{\epsilon }+2 e^{\frac{4 i k}{\sqrt{e} H}+\frac{2 i k}{H \epsilon }+1} H k^3 (2 k+a H i \epsilon )\right)
\ees
\ewt

\bwt
\begin{align}
\mathcal{D} =\,\, &4 a e^{\frac{4 i k}{\sqrt{e} H}+\frac{2 i k}{H \epsilon }+\frac{3}{2}} H^4 k^2 \epsilon ^{3/2}-e^{\frac{2 i \left(\sqrt{\epsilon }+2\right) k}{\sqrt{e} H \sqrt{\epsilon }}+\frac{2 (a-1) i k}{a H \epsilon }+\frac{5}{2}} H^5 (2 i k+a H \epsilon ) \epsilon ^{3/2}+e^{\frac{2 i \left(2 \epsilon +\sqrt{\epsilon }\right) k}{\sqrt{e} H \epsilon }+\frac{2 (a-1) i k}{a H \epsilon }+\frac{5}{2}} H^5 (2 i k+a H \epsilon ) \epsilon ^{3/2}\nn\\
&-a e^{\frac{4 i k}{\sqrt{e} H}+\frac{2 i k}{H \epsilon }+\frac{5}{2}} H^6 \epsilon ^{5/2}+a e^{\frac{2 i \left(\epsilon +\sqrt{\epsilon }\right) k}{\sqrt{e} H \epsilon }+\frac{2 i k}{H \epsilon }+\frac{5}{2}} H^6 \epsilon ^{5/2}-2 i a e^{\frac{2 i \left(\epsilon +\sqrt{\epsilon }\right) k}{\sqrt{e} H \epsilon }+\frac{2 i k}{H \epsilon }+1} H^3 k^3 \epsilon ^2-3 i a e^{\frac{4 i k}{\sqrt{e} H}+\frac{2 i k}{H \epsilon }+2} H^5 k \epsilon ^2\nn\\
&-2 a e^{\frac{2 i \left(\epsilon +\sqrt{\epsilon }\right) k}{\sqrt{e} H \epsilon }+\frac{2 i k}{H \epsilon }+\frac{3}{2}} H^4 k^2 \left(\sqrt{\epsilon }+1\right) \epsilon ^2+a e^{\frac{2 i \left(\epsilon +\sqrt{\epsilon }\right) k}{\sqrt{e} H \epsilon }+\frac{2 i k}{H \epsilon }+2} H^5 i k \left(2 \sqrt{\epsilon }+1\right) \epsilon ^2+2 a e^{\frac{4 i k}{\sqrt{e} H}+\frac{2 i k}{H \epsilon }+1} H^3 i k^3 \epsilon\nn\\
&+e^{\frac{2 i \left(2 \epsilon +\sqrt{\epsilon }\right) k}{\sqrt{e} H \epsilon }+\frac{2 (a-1) i k}{a H \epsilon }+2} H^4 i k (2 i k+a H \epsilon ) \epsilon +2 e^{\frac{2 i \left(\sqrt{\epsilon }+2\right) k}{\sqrt{e} H \sqrt{\epsilon }}+\frac{2 (a-1) i k}{a H \epsilon }+\frac{3}{2}} H^3 k^2 \left(\sqrt{\epsilon }-1\right) (2 i k+a H \epsilon ) \epsilon \nn\\
&+e^{\frac{2 i \left(\sqrt{\epsilon }+2\right) k}{\sqrt{e} H \sqrt{\epsilon }}+\frac{2 (a-1) i k}{a H \epsilon }+2} H^4 k \left(2 \sqrt{\epsilon }-1\right) (2 k-i a H \epsilon ) \epsilon -4 e^{\frac{2 i \left(\sqrt{\epsilon }+2\right) k}{\sqrt{e} H \sqrt{\epsilon }}+\frac{2 (a-1) i k}{a H \epsilon }+\frac{1}{2}} H k^4 (2 i k+a H \epsilon )\nn\\
&+4 e^{\frac{2 i k \left(\sqrt{e} (a-1)+a \left(\epsilon +2 \sqrt{\epsilon }\right)\right)}{a \sqrt{e} H \epsilon }} k^5 (2 k-i a H \epsilon )+2 e^{\frac{2 i \left(\sqrt{\epsilon }+2\right) k}{\sqrt{e} H \sqrt{\epsilon }}+\frac{2 (a-1) i k}{a H \epsilon }+1} H^2 k^3 (\epsilon -1) (2 k-i a H \epsilon)
\end{align}
\ewt

\bibliographystyle{utcaps}

\begin{thebibliography}{10}

\bibitem{schwinger}
J. Schwinger, Phys. Rev. 82, 664 (1951); for a text book: C. Itzykson and J. B. Zuber, Quantum Field Theory (McGraw-Hill, New York, 1980).

\bibitem{hawk} S.W. Hawking, `Particle Creation By Black Holes', Commun. Math. Phys. 43, 199 (1975) [Erratum-ibid. 46, 206 (1976)].

\bibitem{early} L. Parker, Phys. Rev. Lett. 21, 562 (1968); Phys. Rev. 183, 1057 (1969); Phys. Rev. D 3, 346 (1971); L. P. Grishchuk, Sov. Phys. JETP 40, 409 (1975).

\bibitem{books}
N. D. Birrell and P. C. W. Davies, Quantum Fields in Curved Space (Cambridge Univ. Press, Cambridge, 1982); S. A. Fulling, Aspects of Quantum Field Theory in Curved Spacetime (Cambridge Univ. Press, Cambridge, 1989); V. Mukhanov and S. Winitzki, Inroduction to Quantum Effects in Gravity (Cambridge University Press, Cambridge,
2007).

\bibitem{reviews} 
B. S. DeWitt, Phys. Rept. 19, 295 (1975); R. Brout, S. Massar, R. Parentani, P. Spindel, Phys. Rept. 260, 329 (1995); G. V. Dunne, in `From Fields to Strings: Circumnavigating Theoretical Physics', Shifman, M. (ed.) et al. (2004)[arXiv:hep-th/0406216]; T. Padmanabhan, Phys. Rept. 406, 49 (2005) [arXiv:gr-qc/0311036]; D. N. Page, New J. Phys. 7, 203 (2005) [arXiv:hep-th/0409024].

\bibitem{paddy} 
T. Padmanabhan, Pramana, 37, 179 (1991);
L. Sriramkumar and T. Padmanabhan, Phys. Rev. D 54, 7599 (1996); 
K.~Srinivasan and  T.~Padmanabhan, Phys. Rev. D 60, 024007 (1999), [arXiv:gr-qc/9812028v1];
L. Sriramkumar and T. Padmanabhan, Int. J. Mod. Phys. D 11: 1 (2002) [arXiv:gr-qc/9903054].

\bibitem{others}
Y. Kluger et al., Phys. Rev. Lett. 67, 2427 (1991); 
C. Kiefer et al., Class. Quant. Grav. 8, L185 (1991);
C. Kiefer, Phys. Rev. D 45, 2044 (1992); 
A. Campos and E. Verdaguer, Phys. Rev. D 49, 1861 (1994) [arXiv:gr-qc/9307027];
B. L. Hu, G. Kang and A. Matacz, Int. J. Mod. Phys. A9, 991 (1994) [arXiv:gr-qc/9312014];
S. Habib, C. Molina-Paris and E. Mottola, Phys. Rev. D 61, 024010 (1999) [arXiv:gr-qc/9906120];
S. P. Kim, D. N. Page, Phys. Rev. D 73, 065020 (2006); 
F. Cooper and G. C. Nayak (2006) [arXiv:hep-th/0611125]; 
S. P. Kim et al., Phys. Rev. D 75, 045013 (2007).


\bibitem{brand} R.H. Brandenberger, Rev. Mod. Phys. 57, 1 (1985); 
V. F. Mukhanov, H. A. Feldman and R. H. Brandenberger, Phys. Rept. 215, 203 (1992); 
J. Martin, arXiv:0704.3540 [hep-th]; Lect. Notes Phys. 669, 199 (2005) [arXiv: hep-th/0406011].

\bibitem{paddy} T. Padmanabhan, Theoretical Astrophysics, Volume III: Galaxies and Cosmology, (Cambridge University Press, Cambridge, England, 1999), Sec. 5.7;  A. R. Liddle and D. H. Lyth, Cosmological Inflation and Large-Scale Structure (Cambridge University Press, Cambridge, 1999).

\bibitem{othr}S. W. Hawking, Phys. Lett. B 115, 295 (1982); 
A. A. Starobinsky, Phys. Lett. B 117, 175 (1982); A. H. Guth and S.-Y. Pi, Phys. Rev. Lett. 49, 1110 (1982); A. D. Linde, Phys. Lett. B 116, 335 (1982); J. M. Bardeen, P. J. Steinhardt and M. S. Turner, Phys. Rev. D 28, 679 (1983);
 L. F. Abbott and M. B. Wise, Nucl. Phys. B 244, 541 (1984); 
T. Padmanabhan, Phys. Rev. Lett. 60 2229 (1988); 
T. Padmanabhan, T. R. Seshadri and T. P. Singh, Phys. Rev. D 39, 2100 (1989).

\bibitem{gaurang2007}
G.~Mahajan and T.~Padmanabhan, Gen.~Rel.~Grav., 40, 661 (2008), [arXiv:gr-qc/0708.1233];
G.~Mahajan and T.~Padmanabhan, Gen.~Rel.~Grav., 40, 709 (2008), [arXiv:gr-qc/0708.1237].

\bibitem{gaurang2008}
G.~Mahajan, Annals. Phys. 324, 361 (2009). 

\bibitem{sriram}
L. Sriramkumar and T. Padmanabhan, Phys. Rev. D 71, 103512 (2005), [arXiv:gr-qc/0408034]  

\bibitem{cosmin}
T. Padmanabhan,  Res. Astro. Astrophys.,12 , 891 (2012) [arXiv:1207.0505];
Hamsa Padmanabhan, T. Padmanabhan, Int.Jour.Mod.Phys, D 22, 1342001 (2013) [arXiv:1302.3226].    

\bibitem{bunchdavies}
T.~S.~Bunch and P.~C.~W~Davies, J. Phys. A, 11, 1315 (1978).

\bibitem{suprit1302}
S.~Singh, C.~Ganguly and T.~Padmanabhan, Phys. Rev. D 87, 104004 (2013), [arXiv:1302.7177].

\bibitem{lyth2008}
D.~H.~Lyth and D.~Seery, Phys. Lett. B, 662, 309 (2008) [arXiv:astro-ph/0607647]. 

\bibitem{gravitationTP}
T.~Padmanabhan, {\sl Gravitation: Foundations and Frontiers}, Cambridge University Press (2010).

\end{thebibliography}
\providecommand{\href}[2]{#2}\begingroup\raggedright\endgroup
\end{document}